\documentclass[english,preprint,prd,groupedaddress,nofootinbib,showpacs]{revtex4-1}
\usepackage[T1]{fontenc}
\usepackage[utf8]{luainputenc}
\usepackage{color}
\usepackage{amsmath}
\usepackage{amssymb}
\usepackage{esint}
\usepackage[dvips]{graphicx}
\usepackage{pstricks,pst-coil}
\usepackage{pst-all}
\RequirePackage{flushend}
\RequirePackage[colorlinks,citecolor=blue,urlcolor=blue,linkcolor=blue]{hyperref}

\makeatletter

\providecommand{\tabularnewline}{\\}

\@ifundefined{textcolor}{}
{%
 \definecolor{BLACK}{gray}{0}
 \definecolor{WHITE}{gray}{1}
 \definecolor{RED}{rgb}{1,0,0}
 \definecolor{GREEN}{rgb}{0,1,0}
 \definecolor{BLUE}{rgb}{0,0,1}
 \definecolor{CYAN}{cmyk}{1,0,0,0}
 \definecolor{MAGENTA}{cmyk}{0,1,0,0}
 \definecolor{YELLOW}{cmyk}{0,0,1,0}
}

\usepackage[english]{babel}

\usepackage{latexsym}\usepackage{dcolumn}\usepackage{bm}\usepackage{subfigure}

\def\uma{\rm 1\!\!\hskip 1 pt l}

\makeatother

\usepackage{babel}
\begin{document}
\begin{abstract}
We here propose a 5-dimensional  Abelian gauge model based on the
mixing between a $U(1)$  potential and an Abelian 3-form 
field by means of a topological mass term. An extended covariant
derivative is introduced to minimally couple a Dirac field to the
$U(1)$ potential, while this same covariant derivative non-minimally
couples the 3-form field to the charged fermion. A number of properties
are discussed in 5D; in particular, the appearance of a topological
fermionic current. A 4-dimensional reduced version of the model is
investigated and, in addition to the $U(1)$ electric- and magnetic-sort of 
fields, there emerges an extra set of electric- and magnetic-like
fields which contribute a negative pressure and may be identified
as a possible fraction of dark energy. The role of the topological
fermionic current is also contemplated upon dimensional reduction from
5D to 4D. Other issues we present in 4 space-time dimensions
are the emergence  of a pseudo-scalar massive particle,
an extra massive neutral gauge boson, which we
interpret as a kind of paraphoton, and
the calculation of spin- and velocity-dependent interparticle potentials associated to
the exchange of the intermediate bosonic fields of the model.
\end{abstract}


\title{A 3-form Gauge Potential in 5D in connection with a Possible Dark
Sector of 4D-Electrodynamics}



\author{D. Cocuroci}

\email{cocuroci@cbpf.br}

\affiliation{Centro Brasileiro de Pesquisas F\'{i}sicas, Rua Dr. Xavier Sigaud
150, Urca,\\
 Rio de Janeiro, Brazil, CEP 22290-180}

\author{M. J. Neves}

\email{mariojr@ufrrj.br}

\affiliation{Universidade Federal Rural do Rio de Janeiro,\\
BR 465-07, 23890-971, Serop\'edica, Rio de Janeiro, Brazil}

\author{J. A. Helayël-Neto}

\email{helayel@cbpf.br}

\author{L. P. R. Ospedal}

\email{leoopr@cbpf.br}

\affiliation{Centro Brasileiro de Pesquisas F\'{i}sicas, Rua Dr. Xavier Sigaud
150, Urca,\\
 Rio de Janeiro, Brazil, CEP 22290-180}


\maketitle

\section{Introduction\label{intro}}


The possibility of a multidimensional Universe has raised a growing
interest over the past decades. Currently, the reasons for this interest
come primarily from approaches such as Superstring Theory, which is
able to incorporate quantum gravity in a natural and consistent way
\citep{green_superstrings_1999}.

As a consequence of the superstring landscape, it is nowadays widely
accepted that the structure of spacetime must be described as the
product of a 5-dimensional Anti-de Sitter space by a 5-dimensional
hypersphere. Thus, we adopt the viewpoint that the fundamental physics
may be derived from five space-time and five compact internal dimensions.

In addition, the possibility of an equivalence between a classical
gravity theory, defined in a 5-dimensional (spacetime) bulk, and
a quantum gauge theory (Yang-Mills) on the corresponding 4-dimensional
boundary was first proposed by Maldacena in 1997 \citep{maldacena_large_1997}.
Important aspects of the gravity-gauge correspondence were elaborated
in articles by Gubser, Klebanov and Polyakov, and by Edward Witten
\citep{gubser_gauge_1998,witten_anti_1998,aharony_large_2000}.

We shall not however adopt the $AdS_{5}/CFT_{4}$ equivalence in its
full sense. What we borrow from this correspondence is simply the
point of view that our fundamental physics takes place in five space-time
dimensions; whether this physics should be specifically analyzed in
an $AdS_{5}$ or a 5D Minkowski scenario will actually depend on the
particular phenomenon under study. Here, we shall assume that, so
long as the energy scale for electromagnetic interactions is considered,
we do not need to consider the presence of a cosmological constant
in the 5-dimensional world. For the investigation we aim to pursue,
our onset is indeed a 5-dimensional Minkowski space-time.

Actually, in the present study, we explore the consequences of an
extra dimension \citep{arkanihamed_hierarchy_1998}, by just considering
Minkowski space as the background space-time, because the effect of
the curvature of the Anti-de Sitter space (induced by a cosmological
constant, which for the LAMBDA-CDM model is taken to be $~10^{-47}$\,$GeV^{4}$
\citep{tegmark_cosmological_2004}) yields negligible corrections
as compared to the scale of masses and lengths typical of QED processes
\citep{meszaros_high-energy_1992}. By neglecting the cosmological
constant, the isometry group of $AdS_{5}$ (namely, $SO(2,4)$) reduces
to the Poincaré group in five dimensions. So, we shall here consider
a model for electromagnetic interactions in a 5-dimensional Minkowski
space and our 4-dimensional physics must come out as the result of
a specific dimensional reduction scheme rather than by holographic
projection.

It is noteworthy that, if we were considering the quantum effects
of gravitation, the cosmological constant should not be neglected,
for it is known that the latter induces the production of gravitons
with mass of the order of the Planck mass \citep{stelle_renormalization_QG_1977,Sezgin_vanNieuwenhuizen_ghostfree_1980}.
However, in the particular case we are concerned to study, massive
gravitons do not couple to the associated fluctuations of the electron
and photon due to the fact that they are highly massive, so that,
in the energy regime of validity of QED processes, those gravitons
with so a huge mass (induced by the cosmological constant) are not
excited.

 From this perspective, in this paper, we start off from a model
based on the association of a 3-form gauge potential with accelerated
expansion of the Universe \citep{koivisto_inflation_2009}. The introduction
of the concept of dark energy is actually one of the main approaches
to account for the phenomenon of a Universe in accelerated expansion
\citep{peebles_cosmological_2003}. Our particular model, formulated
in five space-time dimensions, as already anticipated above, also
yields, upon a dimensional reduction mechanism, the appearance of
an extra neutral massive boson in 4-dimensional Minkowski space 
\citep{de_blas_combining_2012,dobrich_2014kda,kazanas2015}.
This shall be presented in details in the sequel. We are actually interested in a 
gauge-invariant mass term which plays the role of a 
mixed Chern-Simons topological mass, as it may provide a scenario in four space-time 
dimensions where an axionic particle and a sort of paraphoton emerge together. The 
pseudo-scalar (axion) and the pseudo-vector are unified in the 5-dimensional world 
through a topological mass term.

In connection with the study of the 3-form potential \citep{ovrut_membranes_1997,hitchin_geometry_2000,youm_bogomolnyi-prasad-sommerfield_2001,grana_gauge-gravity_2002,gerasimov_towards_2004,Spallucci_three-form_2004,dvali_three-form_2005,bizdadea_couplings_2006,ho_m5-brane_2008,cioroianu_consistent_2009,jimenez_perturbations_2009,koivisto_inflation_2009-1,koivisto_inflation_2009,koivisto_three-form_2010,ngampitipan_dynamics_2011,koivisto_coupled_2012,schmude_supersymmetry_2012},
the mass of the photon is included in order to seek a situation that
is as broad as possible, i.e., capable of exploring all the possibilities
that a 3-form may offer. According to the work by Koivisto and Nunes
\citep{koivisto_inflation_2009}, 3-forms are used to describe the
dark energy fraction of our presently-expanding Universe. We should
however point out that the 3-form potential was initially studied
by these authors through a kinetic term (minimally coupled to Einstein
gravity) added up to a potential term \citep{koivisto_three-form_2010}.
Subsequently, the 3-form was reassessed to include coupling to point
particles \citep{koivisto_coupled_2012}. Here, we intend to investigate
the 3-form in association with an Abelian gauge vector, in a 5D scenario,
by introducing a topological (mixed) Chern-Simons-like mass term.

On the other hand, in a recent paper \citep{diamantini_gauge_2013},
the authors show how a vortex gauge field, whenever coupled to charged
fermions, induce, by radiative corrections, a gauge invariant mass
term for the photon. Rather than as a dynamical effect, like in the
paper \citep{diamantini_gauge_2013}, in our work, this mass term
arises from a dimensional reduction from the 5D model where there
is a topological mass term, as it is going to be shown in the subsequent
Sections.

Five-dimensional Chern-Simons theory in its Abelian version has recently
been studied by Qi, Witten and Zhang (QWZ) in the context of the physics
of topological superconductors \citep{qi_axion_2013}. We also take
this remarkable contribution - in addition to the $AdS_{5}/CFT_{4}$
correspondence - as a motivation for our exploitations in a 5-dimensional
space-time. As it is known, in superconductivity, a massive photon
must be present to accommodate the Meissner effect, responsible for
the expulsion of the magnetic field from inside materials in their
superconducting phase. Thus, with the physics of topological superconductors
being processed in five dimensions, according to the QWZ scenario,
the photon could acquire mass through a mechanism of topological mass
generation, as we are going to present here.

In summary, we intend to explore an electrodynamic model that uses
both a 3-form and a 1-form gauge potentials in a mixed way in order
to generate a massive gauge boson in a 5-dimensional scenario. Upon
dimensional reduction \citep{scherk_spontaneous_1979}, we actually
attain a model that presents in its spectrum a massive neutral vector
boson degenerated (i.e., with the same mass) with a neutral scalar
excitation,  the latter produced by the mixing between a genuine 
pseudo-scalar field
and a longitudinal vector field. Our work follows the outline below.

In Section \ref{sec:modelo}, we present the model we adopt to pursue
our investigation. We split it in two subsections, in which we obtain
the equation fields, discuss the conservation laws and carry out the
dimensional reduction of the model to 4D. Next, in Section \ref{sec:fermionic},
we add up $U(1)-$charged fermions to the action of the model in 5D discussed in the
previous Section. We obtain the fermionic conserved currents in 5D
and connect them to the pseudo-tensor current of the paper \citep{diamantini_gauge_2013}.
Coupling the model to gravity in 5D is also carried out. The 5-dimensional action is
then reduced to 4D, and we calculate the propagators
of the bosonic sector to read off the spectrum of excitations. As an application,
interparticle potentials generated by the intermediation of the bosonic fields exchanged
by external currents are worked out. Finally,
our Concluding Comments are cast in Section \ref{sec:Concluding}.


\section{Description of the Model}

\label{sec:modelo}


Taking for granted the importance of understanding physics in our
4-dimensional world from a more fundamental 5-dimensional physics,
we focus here on a study of a specific electrodynamic model in 5 dimensions
aiming at the possible consequences it yields in a 4-dimensional space-time.

Thus, in this Section, we present the model which consists of a Lagrangian
density containing the kinetic terms for each gauge field ($A_{\bar{\mu}}$,
and $C_{\bar{\mu}\bar{\nu}\bar{\kappa}}$), and a mixing term between
them. This mixing term is capable of ensuring that the mass of the
associated particle is independent of the metric characteristics of
the space. It is known in the literature as a topological term \citep{abbott_background_1983,christiansen_n_1999,ferreira_cosmic_2000}.
We also exhibit the field equations, the Bianchi identities and the
conservation laws.

Consider the action in 5D whose corresponding Lagrangian density is
as follows:
\begin{equation}
\mathcal{L}=-\frac{1}{4}\,F_{\bar{\mu}\bar{\nu}}F^{\bar{\mu}\bar{\nu}}
+\alpha \, H_{\bar{\mu}\bar{\nu}\bar{\kappa}\bar{\lambda}}H^{\bar{\mu}\bar{\nu}\bar{\kappa}\bar{\lambda}}
+\beta \, \epsilon^{\bar{\mu}\bar{\nu}\bar{\kappa}\bar{\lambda}\bar{\rho}}A_{\bar{\mu}}\partial_{\bar{\nu}}
C_{\bar{\kappa}\bar{\lambda}\bar{\rho}}\;,\label{eq:lagrangeano5D}
\end{equation}
where $A_{\bar{\mu}}$ is the  Abelian vector and $C_{\bar{\mu}\bar{\nu}\bar{\kappa}}$
is the 3-form gauge  field, one of the main elements of this study. The notation
of the indices in 5 dimensions is $\bar{\mu}=\{0,1,2,3,4\}.$ 
The tensor $F_{\bar{\mu}\bar{\nu}}$ is the usual electromagnetic
field strength, and the tensor $H_{\bar{\mu}\bar{\nu}\bar{\kappa}\bar{\lambda}}$
is the completely antisymmetric field strength associated with the
3-form field, $C_{\bar{\mu}\bar{\nu}\bar{\kappa}}$:
\begin{equation}
H_{\bar{\mu}\bar{\nu}\bar{\kappa}\bar{\lambda}}=\partial_{\bar{\mu}}C_{\bar{\nu}\bar{\kappa}\bar{\lambda}}
-\partial_{\bar{\nu}}C_{\bar{\lambda}\bar{\kappa}\bar{\mu}}+\partial_{\bar{\kappa}}C_{\bar{\lambda}\bar{\mu}\bar{\nu}}
-\partial_{\bar{\lambda}}C_{\bar{\mu}\bar{\nu}\bar{\kappa}}\;.\label{eq:tensorH}
\end{equation}
The parameters $\alpha$ and $\beta$ are both real. It is not dificult
to check that the $\beta$-parameter has mass dimension. The action
defined through the Lagrangian (\ref{eq:lagrangeano5D}) is invariant
under the following Abelian gauge transformations in 5D:
\begin{equation}
A_{\bar{\mu}}\longmapsto A_{\bar{\mu}}^{\;\prime}=A_{\bar{\mu}}+\partial_{\bar{\mu}}\Lambda\,\;,\ \label{eq:Transf_Gauge_A_mu}
\end{equation}
\begin{equation}
C_{\bar{\mu}\bar{\nu}\bar{\kappa}}\longmapsto C_{\bar{\mu}\bar{\nu}\bar{\kappa}}^{\;\prime}=C_{\bar{\mu}\bar{\nu}\bar{\kappa}}+\partial_{\bar{\mu}}\xi_{\bar{\nu}\bar{\kappa}}+\partial_{\nu}\xi_{\bar{\kappa}\bar{\mu}}+\partial_{\bar{\kappa}}\xi_{\bar{\mu}\bar{\nu}}\;,\label{eq:Transf_Gauge_C_mu_nu_kappa}
\end{equation}
where $\Lambda$ and $\xi_{\bar{\mu}\bar{\nu}}$ are real functions
and $\xi_{\bar{\mu}\bar{\nu}}$ is a antisymmetric tensor field. The
transformation (\ref{eq:Transf_Gauge_A_mu}) is the already known
one from electrodynamics, $U(1)_{A_{\bar{\mu}}}$, whereas (\ref{eq:Transf_Gauge_C_mu_nu_kappa})
is the antisymmetrized version of the gauge transformation for a rank-3
tensor, $U(1)_{C_{\bar{\mu}\bar{\nu}\bar{\kappa}}}$. Thus, the action
is said to be $U(1)_{A_{\bar{\mu}}}\otimes U(1)_{C_{\bar{\mu}\bar{\nu}\bar{\kappa}}}$-invariant.
The Lagrangian (\ref{eq:lagrangeano5D}) gives us the field equations
\begin{equation}
\partial_{\bar{\mu}}F^{\bar{\mu}\bar{\nu}}+6\beta\widetilde{H}^{\bar{\nu}}=0\;,\ \label{eq:Eqs_Campo_a}
\end{equation}
\begin{equation}
8\alpha\;\partial_{\bar{\mu}}H^{\bar{\mu}\bar{\nu}\bar{\kappa}\bar{\lambda}}-\beta\widetilde{F}^{\bar{\nu}\bar{\kappa}\bar{\lambda}}=0\;.\ \label{eq:Eqs_Campo_b}
\end{equation}
where the relations between the dual tensors $\widetilde{F}^{\bar{\mu}\bar{\nu}\bar{\kappa}}$
and $F_{\bar{\mu}\bar{\nu}}$ are given by the expressions
\begin{equation}
F_{\bar{\mu}\bar{\nu}}=-\frac{1}{3!}\,\epsilon_{\bar{\mu}\bar{\nu}\bar{\kappa}\bar{\alpha}\bar{\beta}}\widetilde{F}^{\bar{\mu}\bar{\nu}\bar{\kappa}}
\hspace{0.4cm} \mbox{and} \hspace{0.4cm} \widetilde{F}_{\bar{\mu}\bar{\nu}\bar{\kappa}}=\frac{1}{2!}\,\epsilon_{\bar{\mu}\bar{\nu}\bar{\kappa}\bar{\lambda}\bar{\rho}}F^{\bar{\lambda}\bar{\rho}}\;.\ \label{eq:Dual_tensors}
\end{equation}
As for the relations between $\widetilde{H}^{\bar{\mu}}$ and $H_{\bar{\mu}\bar{\nu}\bar{\kappa}\bar{\lambda}}$,
the expressions are given by
\begin{equation}
H_{\bar{\mu}\bar{\nu}\bar{\kappa}\bar{\lambda}}=\epsilon_{\bar{\mu}\bar{\nu}\bar{\kappa}\bar{\lambda}\bar{\rho}}\widetilde{H}^{\bar{\rho}}
\hspace{0.4cm} \mbox{and} \hspace{0.4cm} \widetilde{H}_{\bar{\mu}}=\frac{1}{4!}\,\epsilon_{\bar{\mu}\bar{\nu}\bar{\kappa}\bar{\lambda}\bar{\rho}}H^{\bar{\nu}\bar{\kappa}\bar{\lambda}\bar{\rho}}\;.\label{eq:Dual_tensors-1}
\end{equation}
The Bianchi identities associated to the fields $F^{\bar{\mu}\bar{\nu}}$
and $H_{\bar{\mu}\bar{\nu}\bar{\kappa}\bar{\lambda}}$ are, respectively:
\begin{equation}
\partial_{\bar{\mu}}F_{\bar{\nu}\bar{\kappa}}+\partial_{\bar{\nu}}F_{\bar{\mu}\bar{\kappa}}+\partial_{\bar{\kappa}}F_{\bar{\mu}\bar{\nu}}=0\;,\ \label{eq:ID_Bianchi5D_F}
\end{equation}
\begin{eqnarray}
\partial_{\bar{\mu}}H_{\bar{\nu}\bar{\kappa}\bar{\lambda}\bar{\rho}}+\partial_{\bar{\nu}}H_{\bar{\kappa}\bar{\lambda}\bar{\rho}\bar{\mu}}+\partial_{\bar{\kappa}}H_{\bar{\nu}\bar{\lambda}\bar{\mu}\bar{\rho}}
+\partial_{\bar{\lambda}}H_{\bar{\nu}\bar{\kappa}\bar{\rho}\bar{\mu}}+\partial_{\bar{\rho}}H_{\bar{\mu}\nu\bar{\kappa}\bar{\lambda}}=0\;.\label{eq:ID_Bianchi5D_H}
\end{eqnarray}
Expression (\ref{eq:ID_Bianchi5D_H}) can also be cast in a more compact
form in terms of the dual of $H_{\bar{\mu}\bar{\nu}\bar{\kappa}\bar{\lambda}}$,
i.e.: 
\begin{equation}
\partial_{\bar{\mu}}\widetilde{H}^{\bar{\mu}}=0\;.\label{eq:Bianchi_compacta}
\end{equation}
The field equations (\ref{eq:Eqs_Campo_a}) and (\ref{eq:Eqs_Campo_b})
are coupled and we must necessarily decouple them in order to implement
the procedure that will reveal the mass of the particle(s) associated(s)
with both fields. It yields :
%
\begin{equation}
\left(\Box-\frac{3}{4\alpha}\beta^{2}\right)F_{\bar{\mu}\bar{\nu}}=0\,,\label{eq:Field_eq_mass-a}
\end{equation}
and 
\begin{equation}
\left(\Box-\frac{3}{4\alpha}\beta^{2}\right)H_{\bar{\mu}\bar{\nu}\bar{\kappa}\bar{\lambda}}=0\;.\label{eq:Field_eq_mass-b}
\end{equation}
Therefore, it is noted from (\ref{eq:Field_eq_mass-a}-\ref{eq:Field_eq_mass-b})
both fields are shown to exhibit the same mass term, which is given by
$\xi=\frac{-3\beta^{2}}{4\alpha}$,
where, it is considered that the parameter $\alpha$ must be restricted
to a negative real number.
The energy-momentum tensor is obtained by multiplying the eq. (\ref{eq:Eqs_Campo_a})
by $F_{\bar{\nu}\bar{\alpha}}$ and using the following relation between
the dual fields
\begin{equation}
\widetilde{F}^{\bar{\nu}\bar{\kappa}\bar{\lambda}}H_{\bar{\nu}\bar{\kappa}\bar{\lambda}\bar{\alpha}}=-6F_{\bar{\nu}\bar{\alpha}}\widetilde{H}^{\bar{\nu}}\;.\label{eq:Relation_Duals}
\end{equation}
Then, we insert (\ref{eq:Eqs_Campo_b}) to get
%
\begin{eqnarray}
-\partial_{\bar{\mu}}(8\bar{\alpha}H^{\bar{\mu}\bar{\nu}\bar{\kappa}\bar{\lambda}}H_{\bar{\nu}\bar{\kappa}\bar{\lambda}\bar{\alpha}})+8\alpha H^{\bar{\mu}\bar{\nu}\bar{\kappa}\bar{\lambda}}\partial_{\bar{\mu}}H_{\bar{\nu}\bar{\kappa}\bar{\lambda}\bar{\alpha}}
+\partial_{\bar{\mu}}(F^{\bar{\mu}\bar{\nu}}F_{\bar{\nu}\bar{\alpha}})-F^{\bar{\mu}\bar{\nu}}\partial_{\bar{\mu}}F_{\bar{\nu}\bar{\alpha}}=0\;.
\end{eqnarray}
Thus, replacing the Bianchi identities (\ref{eq:ID_Bianchi5D_F}-\ref{eq:ID_Bianchi5D_H})
and using the relation
\begin{eqnarray}
2F_{\bar{\nu}\bar{\alpha}}\partial_{\bar{\mu}}F^{\bar{\mu}\bar{\nu}}=\partial_{\bar{\mu}}(16\alpha H^{\bar{\mu}\bar{\nu}\bar{\kappa}\bar{\lambda}}H_{\bar{\nu}\bar{\kappa}\bar{\lambda}\bar{\alpha}})+\partial_{\bar{\alpha}}(2\alpha H_{\bar{\mu}\bar{\nu}\bar{\kappa}\bar{\lambda}}^{2})\,,\hspace{0.7cm}\label{relFH}
\end{eqnarray}
we obtain the continuity equation
\begin{equation}
\partial_{\bar{\mu}}\Theta_{\;\;\bar{\alpha}}^{\bar{\mu}}=0 \; , \label{eq:Conservation_law}
\end{equation}
where $\Theta_{\;\;\bar{\alpha}}^{\bar{\mu}}$, the energy-momentum
tensor associated with the Lagrangian (\ref{eq:lagrangeano5D}), is
given by
\begin{eqnarray}
\Theta_{\;\;\bar{\alpha}}^{\bar{\mu}}=-8\alpha\left(H^{\bar{\mu}\bar{\nu}\bar{\kappa}\bar{\lambda}}H_{\bar{\nu}\bar{\kappa}\bar{\lambda}\bar{\alpha}}+\delta_{\;\;\bar{\alpha}}^{\bar{\mu}}\;\frac{1}{8}H_{\bar{\mu}\bar{\nu}\bar{\kappa}\bar{\lambda}}^{\;2}\right)
+F^{\bar{\mu}\bar{\nu}}F_{\bar{\nu}\bar{\alpha}}+\delta_{\;\;\bar{\alpha}}^{\bar{\mu}}\;\frac{1}{4}F_{\bar{\mu}\bar{\nu}}^{\;2}\;.\label{eq:tensor_En_Momentum}
\end{eqnarray}
Comparing the second term of $\Theta_{\:\bar{\alpha}}^{\bar{\mu}}$
with the kinetic term of the rank-3 tensor field in (\ref{eq:lagrangeano5D}),
we can set the value of the parameter as $\alpha=-1/8$. Thus, we
rewrite the mass as $m^{2}:=\frac{-3\beta^{2}}{4\alpha}$ and thus
the value of $\beta$ is fixed. Therefore, the topological mass term,
$\Delta$, is given by
\begin{equation}
\Delta=\frac{m}{\sqrt{6}}\,\epsilon^{\bar{\mu}\bar{\nu}\bar{\kappa}\bar{\lambda}\bar{\rho}}A_{\bar{\mu}}\partial_{\bar{\nu}}C_{\bar{\kappa}\bar{\lambda}\bar{\rho}}\;.\label{eq:massa_topol}
\end{equation}
The energy-momentum tensor is written in terms of the field-
strength tensors, then it is naturally invariant under the gauge transformations
(\ref{eq:Transf_Gauge_A_mu}) and (\ref{eq:Transf_Gauge_C_mu_nu_kappa}).
It is also symmetrical the expression (\ref{eq:tensor_En_Momentum})
can be rewritten in terms of the dual field of $H_{\bar{\mu}\bar{\nu}\bar{\kappa}\bar{\lambda}}$
as below
\begin{equation}
\Theta_{\;\;\bar{\alpha}}^{\bar{\mu}}=6\widetilde{H}^{\bar{\mu}}\widetilde{H}_{\bar{\alpha}}
-\delta_{\;\;\bar{\alpha}}^{\bar{\mu}}\;3\;\widetilde{H}_{\bar{\mu}}^{\;2}+F^{\bar{\mu}\bar{\nu}}F_{\bar{\nu}\bar{\alpha}}
+\delta_{\;\;\bar{\alpha}}^{\bar{\mu}}\;\frac{1}{4}F_{\bar{\mu}\bar{\nu}}^{\;2}\;.\label{eq:tensor_En_Momentum_alpha}
\end{equation}

\subsection{Decomposition into irreducible components of $SO(3)$.}

\label{sub:Decomposition} 

To carry out the decomposition of the energy-momentum tensor (\ref{eq:tensor_En_Momentum_alpha}),
the field equations (\ref{eq:Eqs_Campo_a}-\ref{eq:Eqs_Campo_b})
and the Bianchi identities (\ref{eq:ID_Bianchi5D_F}) and (\ref{eq:Bianchi_compacta})
in terms of irreducible components of $SO(3)$, we initially make
the identification of each sector of $F^{\mu\nu}$ and $\widetilde{H}^{\mu}$
with the corresponding irreducible components of $SO(3)$ as listed
in Table \ref{tab:componentes}: 
\begin{center}
\begin{table}[h]
\centering{}%
\begin{tabular}{ll}
\hline
 & \tabularnewline
\hspace{0.5cm}\hspace{0.2cm}$F^{\bar{\mu}\bar{\nu}}$  & \hspace{0.5cm}\hspace{0.2cm}$\widetilde{H}^{\bar{\mu}}$\tabularnewline
 & \tabularnewline
\hline
 & \tabularnewline
$F^{0i}=-\overrightarrow{E_{i}}$ \hspace{0.5cm}  & \hspace{0.5cm} $\widetilde{H}^{0}=\chi$\tabularnewline
 & \tabularnewline
$F_{ij}=-\epsilon_{ijk}\overrightarrow{B_{k}}$ \hspace{0.5cm}  & \hspace{0.5cm} $\widetilde{H}^{i}=\overrightarrow{Y_{i}}$\tabularnewline
 & \tabularnewline
$F^{04}=-b$ \hspace{0.5cm}  & \hspace{0.5cm} $\widetilde{H}^{4}=S$\tabularnewline
 & \tabularnewline
$F^{i4}=\overrightarrow{e_{i}}$  & \tabularnewline
 & \tabularnewline
\hline
\end{tabular}\caption{Components of tensor field $F^{\bar{\mu}\bar{\nu}}$ and dual tensor
of $H^{\bar{\mu}\bar{\nu}\bar{\rho}\bar{\lambda}}$. The indices i,
j, k = 1, 2, 3 refer to the space components in 4 dimensional space-time.}
\label{tab:componentes}
\end{table}

\par\end{center}

From (\ref{eq:Conservation_law}), we extract the components of the
conserved energy-momentum tensor $\Theta_{\:\bar{\alpha}}^{0}$, so
that the energy, the Poynting vector, and a new density pressure associated
with the extra dimension are expressed respectively as follows: 
\begin{eqnarray}
\Theta_{\;\;0}^{0} & = & \frac{1}{2}(E^{2}+B^{2}+b^{2}+e^{2})-3(\chi^{2}+Y^{2}+S^{2})\hspace{0.7cm}\label{eq:componentes_temporais_a}\\
\Theta_{\;\; i}^{0} & = & -(\overrightarrow{E}\times\overrightarrow{B})_{i}+b\overrightarrow{e_{i}}+6\chi\overrightarrow{Y_{i}}\label{eq:componentes_temporais_b}\\
\Theta_{\;\;4}^{0} & = & -\overrightarrow{E}\cdot\overrightarrow{e}+6\chi S\;.\label{eq:componentes_temporais_c}
\end{eqnarray}
Going on with the procedure for extracting the components of the energy-momentum
tensor, we have that the stress tensors read as below :
\begin{eqnarray}
\Theta_{ij} & = & 
- \overrightarrow{E_{i}}\overrightarrow{E_{j}}- 
\overrightarrow{B_{i}}\overrightarrow{B_{j}}+
\overrightarrow{e_{i}}\overrightarrow{e_{j}}-6\overrightarrow{Y_{i}}\overrightarrow{Y_{j}}\label{eq:componentes_espaciais_a}\\
\Theta_{i4} & = & 
(b\overrightarrow{E}+\overrightarrow{e}\times\overrightarrow{B})_{i}-6\overrightarrow{Y_{i}}S\label{eq:componentes_espaciais_b}\\
\Theta_{44} & = & 
\frac{1}{2}(E^{2}-B^{2}+e^{2}-b^{2})-3(S^{2}+\chi^{2}-Y^{2})\;.\hspace{0.7cm}\label{eq:componentes_espaciais_c}
\end{eqnarray}
In Table \ref{tab:componentes}, $\overrightarrow{E},\overrightarrow{B}$,
$\chi$ and $\overrightarrow{Y}$ are the field strengths associated
to the Maxwell-like field and the 2-form potential, respectively. On the 
other hand, $\overrightarrow{e}$, $b$ and $S$ constitute what we call the dark sector of 
our extended
4-dimensional electrodynamics \citep{de_blas_combining_2012}. 
 We name it dark sector because it is connected to the 3-form potential whose gauge 
symmetry is not associated to any sort of matter charge, contrary to the $U(1)-$symmetry 
of vector bosons whose corresponding charge appears in phase symmetry transfomations. At 
this point, we would like to point
out the work of Reference \citep{Holdom1986}, where the author introduces
a second photon, which he refers to as the shadow photon or paraphoton,
a unobserved photon. In our case, what we dub as the dark sector
is the particle associated to the propagation of $\overrightarrow{e}$
and $b$. In our model, there remains a scalar, $S$, which is also
part of what we call the dark sector. It would be interesting, but
we are not doing this here, if we later work out astrophysical constraints
on this dark sector as, it is done in the series of papers quoted in
References \citep{Holdom:1990xp,Davidson_small_charges_1991,Peskin_Davidson_paraphoton_1994}.

The right-hand side of Einstein's equation is essentially described
by the energy-momentum tensor. This constitutes a unified relation
(arising from the space-time symmetry) between the energy density
and the pressure in the system. In a five-dimensional model, one identifies
in the energy-momentum tensor the presence of a sector able to submit
the system, through a particular configuration of the fields (\ref{eq:componentes_espaciais_c}),
to a negative pressure which, in its turn, characterizes the effect
of accelerated inflation of the Universe, the effect of the so-called
dark energy. As a result of the observations, the inflationary profile
of the Universe changes over time \citep{riess_farthest_2001,turner_sne_2002}.
Currently, it presents itself as accelerated \citep{perlmutter_measurements_1999,riess_observational_1998}.
This changing behavior in the inflationary profile may be the result
of changes in the configuration of the present fields in each phase
of the history of the Universe.

In the paper of Reference \citep{koivisto_inflation_2009}, the author
argues that the tiny value of the cosmological constant can be phenomenologically
explained by the use of a 3-form. We also adopt the 3-form, but we
consider that, for the sake of electromagnetic effects, the cosmological
constant is tiny enough, so that we neglect the curvature of the (Anti-de
Sitter) space. In view of that, we adopt Minkowski space as the spacetime
background. Then, we attribute to the presence of a specific sector
of the energy-momentum tensor in 5D, the effect that mimics dark energy,
by virtue of the use of a 3-form in our model.

From what we have discussed above, our work sets out as a possible
theoretical support to the paper \citep{koivisto_inflation_2009}
in order to provide a justification to the fact that the 3-form potential
yields a negative pressure, as suggested by the presence of the $\Theta_{\;\;4}^{4}$-component
of the energy-momentum tensor (\ref{eq:componentes_espaciais_c}),
which may become negative depending on the particular configuration
of the fields ($\overrightarrow{E}$ ,$\overrightarrow{B}$, $\overrightarrow{e}$
, $b$, $\chi$, $S$ and $\overrightarrow{Y}$).

The topological mass term (\ref{eq:massa_topol}) used in our action
does not affect - by construction - the energy-momentum tensor (\ref{eq:tensor_En_Momentum_alpha}),
once it is metric-independent. Hence, if the $\Theta_{\;\;4}^{4}$-component
shows up as a negative contribution, it happens regardless the mass-like
term we adopt. This $\Theta_{\;\;4}^{4}$, which is negative in 5D,
may play the role of the negative pressure associated to a (positive)
cosmological constant in 4D, which is a possible landscape to support
an accelerated expansion of our Universe.



\subsection{Radiation fields in 4D.}

\label{sub:Radiationfields}

Next we exhibit the field equations in 5D extracted from the Lagrangian
(\ref{eq:lagrangeano5D}) where it is considered fixed constants $\alpha$
and $\beta$ as has been detailed in the previous section. The equations
are expressed in terms of the components $\overrightarrow{E}$, $\overrightarrow{B}$,
$\overrightarrow{e}$ and $b$ of $F^{\mu\nu}$and of the components
$\chi$, $S$ and $\overrightarrow{Y}$ of $H_{\mu\nu\kappa\lambda}$
including the mass terms.

We will adopt a dimensional reduction scheme known as Scherk-Schwarz
reduction \citep{scherk_spontaneous_1979} where it is considered
that all potentials and fields do not depend on the extra dimension,
i.e., it is considered that the derivatives of any field to the fifth
coordinate is null, i.e., $\partial_{4}\left(\mbox{any}\,\mbox{field}\right)=0\,.$
The equation (\ref{eq:Eqs_Campo_a}) in the presence of an external
source $J^{\mu}$, when it is decomposed reveals the following equations:
\begin{equation}
\overrightarrow{\nabla}\cdot\overrightarrow{E}+m\sqrt{6}\chi=\rho\label{gauss}
\end{equation}
\begin{equation}
\overrightarrow{\nabla}\times\overrightarrow{B}+m\sqrt{6}\,\overrightarrow{Y}=\overrightarrow{j}+\frac{\partial\overrightarrow{E}}{\partial t}\label{maxweel01}
\end{equation}
\begin{equation}
\overrightarrow{\nabla}\cdot\overrightarrow{e}+m\sqrt{6}\, S=j_{s}+\frac{\partial b}{\partial t}\;.\label{maxweel02}
\end{equation}
When the equation (\ref{eq:Eqs_Campo_b}) is decomposed, it reveals the equations listed
on Table \ref{tab:tabela_02}: \


\begin{center}
\begin{table}[h]
\centering{}%
\begin{tabular}{ll}
\hline
 & \tabularnewline
$\partial_{\mu}H^{\bar{\mu}\bar{\nu}\bar{\kappa}\bar{\lambda}}+\frac{m}{\sqrt{6}}\widetilde{F}^{\bar{\nu}\bar{\kappa}\bar{\lambda}}=J^{\bar{\nu}\bar{\kappa}\bar{\lambda}}\hspace{0.8cm}$  & $\hspace{0.8cm}J^{\bar{\nu}\bar{\kappa}\bar{\lambda}}$\tabularnewline
 & \tabularnewline
\hline
 & \tabularnewline
$-\overrightarrow{\nabla}S+\frac{m}{\sqrt{6}}(\overrightarrow{e})=\overrightarrow{\lambda}$  & $\lambda_{k}=\frac{1}{2}\epsilon_{ijk}J^{0ij}$\tabularnewline
 & \tabularnewline
$\overrightarrow{\nabla}\times\overrightarrow{Y}+\frac{m}{\sqrt{6}}\overrightarrow{B}=\overrightarrow{\zeta}$  & $\mbox{\ensuremath{\zeta^{i}}}=J^{0i4}$\tabularnewline
 & \tabularnewline
$\frac{\partial\overrightarrow{Y}}{\partial t}+\overrightarrow{\nabla}\chi+\frac{m}{\sqrt{6}}\overrightarrow{E}=\overrightarrow{\sigma}$  & $\mbox{\ensuremath{\sigma_{i}}}=\frac{1}{2}\epsilon_{ijk}J^{jk4}$\tabularnewline
 & \tabularnewline
$\frac{\partial S}{\partial t}+\frac{m}{\sqrt{6}}b=\tau$  & $\tau=-\epsilon_{ijk}J^{ijk}$\tabularnewline
 & \tabularnewline
\hline
\end{tabular}\caption{Field equations and their sources.}
\label{tab:tabela_02}
\end{table}
\par\end{center}

As for the Bianchi identity (\ref{eq:ID_Bianchi5D_F}), when it is
decomposed reveals:
\begin{equation}
\overrightarrow{\nabla}\times\overrightarrow{E}=-\frac{\partial\overrightarrow{B}}{\partial t}\;,
\end{equation}
\begin{equation}
\overrightarrow{\nabla}\cdot\overrightarrow{B}=0\;,
\end{equation}
\begin{equation}
\overrightarrow{\nabla}\times\overrightarrow{e}=0\;,
\end{equation}
\begin{equation}
\overrightarrow{\nabla}b=\frac{\partial\overrightarrow{e}}{\partial t}\;.
\end{equation}
And finally, the second Bianchi identity gives us just one expression:
\begin{equation}
\frac{\partial\chi}{\partial t}+\overrightarrow{\nabla}\cdot\overrightarrow{Y}=0\;.\label{eq:decomp_bianchi_H}
\end{equation}
This is a continuity equation involving the components $(\chi,\overrightarrow{Y})$.
It appoints that
\begin{eqnarray}
\Xi:=\int_{{\cal R}}d^{3}{\bf x}\,\,\chi({\bf x},t)\label{IntChi}
\end{eqnarray}
is a conserved quantity of model.

It is important to clarify that, although we write down and study
Maxwell's equations in the 5 dimensions, we shall actually carry out
a dimensional reduction to (1+3)D and, whenever we consider our electromagnetic
fields confined to the 4-dimensional space, there appear extra fields
which are inherited from 5 dimensions upon our dimensional reduction.
So, we are truly considering our electromagnetic interaction in (1+3)D,
but we take into account new fields that show up as a by-product of
the 5-dimensional space-time where we have set up our physical scenario.



\section{The fermion sector in 5D and its dimensional reduction to 4D.}

\label{sec:fermionic}

In this Section, we add to the action corresponding to (\ref{eq:lagrangeano5D})
a fermion sector in $5$ dimensions:
\begin{equation}\label{acao5D}
S_{5D}=\int d^{5}x \left[\bar{\psi}\;(i\gamma^{\bar{\mu}}D_{\bar{\mu}}-m_{f})\;\psi-\frac{1}{4}F_{\bar{\mu}\bar{\nu}}F^{\bar{\mu}\bar{\nu}}
-\frac{1}{8}H_{\bar{\mu}\bar{\nu}\bar{\kappa}\bar{\lambda}}H^{\bar{\mu}\bar{\nu}\bar{\kappa}\bar{\lambda}}
+\frac{m}{\sqrt{6}}\,\epsilon^{\bar{\mu}\bar{\nu}\bar{\kappa}\bar{\lambda}\bar{\rho}}
A_{\bar{\mu}}\partial_{\bar{\nu}}C_{\bar{\kappa}\bar{\lambda}\bar{\rho}}\right] \; ,
\end{equation}
where we insert the covariant derivative in order to study the interaction
of the Dirac field with the gauge fields
\begin{equation}
D_{\bar{\mu}}:=\partial_{\bar{\mu}}+ieA_{\bar{\mu}}+ig\widetilde{H}_{\bar{\mu}}\;,\label{eq:Derivada_Covariante_5D}
\end{equation}
and the spinor $\psi$ is a Dirac fermionic field in 5D. The $\gamma$-matrices
are defined as $\gamma^{\bar{\mu}}=(\gamma^{\mu},\gamma^{4})$, with
$\gamma^{4}=i\gamma_{5}$ and $\gamma_{5}=i\gamma^{0}\gamma^{1}\gamma^{2}\gamma^{3}$
such that they satisfy the anti-commutation relations
\begin{equation}
\{\gamma^{\mu},\gamma^{\nu}\}=2\eta^{\mu\nu}\hspace{0.2cm},\hspace{0.2cm}\{\gamma^{\mu},\gamma_{5}\}=0\;,\label{eq:anticomutation_relation}
\end{equation}
and the conditions $\left(\gamma_{5}\right)^{\dagger}=\gamma_{5}$,
and $\left(\gamma_{5}\right)^{2}=1$.

As already stated previously, the fermionic matter is charged only under the 
$U(1)-$symmetry of the vector field. It has no charge under the Abelian symmetry of the 
3-form gauge potential; this is why the latter is only non-minimally coupled to the 
3-form $C_{\mu \nu \kappa}$-field.

The field equations derived from for the gauge fields in the presence
of fermions are given by
\begin{equation}
\partial_{\bar{\mu}}F^{\bar{\mu}\bar{\nu}}+\sqrt{6}m\widetilde{H}^{\bar{\nu}}=e\bar{\psi}\gamma^{\bar{\nu}}\psi\;,\label{eq:Gauge_fermion_a}
\end{equation}
and
\begin{equation}
\partial_{\bar{\mu}}H^{\bar{\mu}\bar{\nu}\bar{\kappa}\bar{\lambda}}+\frac{m}{\sqrt{6}}\widetilde{F}^{\bar{\nu}\bar{\kappa}\bar{\lambda}}
=4g\epsilon^{\bar{\mu}\bar{\rho}\bar{\nu}\bar{\kappa}\bar{\lambda}}\partial_{\bar{\mu}}(\bar{\psi}\gamma_{\bar{\rho}}\psi)\;.\label{eq:Gauge_fermion_b}
\end{equation}
from which we identify the source terms for each equation:
\begin{equation}
J_{\; F}^{\bar{\mu}}=e\bar{\psi}\gamma^{\bar{\mu}}\psi\label{eq:JF}
\end{equation}
and
\begin{equation}
J_{H}^{\bar{\mu}\bar{\nu}\bar{\kappa}}=4g{\color{black}\epsilon^{\bar{\mu}\bar{\nu}\bar{\kappa}\bar{\lambda}\bar{\rho}}{\color{red}{\color{black}\partial}_{{\color{black}\bar{\lambda}}}}}(\bar{\psi}\gamma_{\bar{\rho}}\psi)\;.\label{eq:JFJH}
\end{equation}
We may notice that these currents arise due to the presence of the
mixing term between the gauge fields in the Lagrangian. $J_{H}^{\bar{\mu}\bar{\nu}\bar{\kappa}}$
is a topological current, which means that we have a current that
is conserved without any reference to the equations of motion and
no continuous symmetry of the Lagrangian or the action is associated
to this conservation equation. In other words, we have an identically
conserved current.

The current $J_{H}^{\bar{\mu}\bar{\nu}\bar{\kappa}}$ above, when
dimensionally reduced to 4D, gives rise precisely to the pseudo-tensor
current to which the vortex gauge field of \citep{diamantini_gauge_2013}
couples. In our case, the current stems from the non-minimal coupling
present in the covariant derivative (\ref{eq:Derivada_Covariante_5D})
as an imprint of the five-dimensional world. So, this topological
current in 5D plays the crucial role of inducing the gauge invariant
mass term of reference \citep{diamantini_gauge_2013} upon its coupling
to the vortex gauge field.


\subsection{Dimensional reduction.}

\label{sub:Dimensional-reduction}

Next, one redefines the complete action, but now having undergone
a procedure of dimensional reduction from five to four dimensions.
The Greek indices follow the notation $\bar{\mu}=(\mu,4)$ where $\mu$
indicates the usual four dimensions and $\bar{\mu}$ indicate five
dimensions, i.e., the four usual dimensions plus a extra spatial dimension.

Here, the 1-form $A^{\bar{\mu}}$ can be divided into a vector sector
and a scalar sector: $A^{\bar{\mu}}=(A^{\mu},A^{4}).$ As for the
3-form, it can be split into two tensor sectors $C^{\bar{\mu}\bar{\nu}\bar{\kappa}}=(C^{\mu\nu\kappa},C^{\mu\nu4})$.
One redefines the scalar component as $A^{4}=\phi$ and one then identifies
the sector $C^{\mu\nu4}=\frac{1}{\sqrt{3}}B^{\mu\nu}$ as the one
known in the literature as the Kalb-Ramond field \citep{kalb_classical_1974}.

Thus, the $5D$ action is reduced to 4D and can be expressed as follows:
\begin{eqnarray}\label{S4D}
S_{4D} & = & \left.\int d^{4}x\left[\bar{\psi}\;(i\gamma^{\mu}D_{\mu}-m_{f})\;\psi-\frac{1}{4}\,F_{\mu\nu}^{\;2}+\frac{1}{6}\,G_{\mu\nu\kappa}^{\;2}
-\frac{2\sqrt{2}}{3}\,m\,\epsilon^{\mu\nu\kappa\lambda}A_{\mu}\partial_{\nu}B_{\kappa\lambda}\right.\right.
\nonumber \\
&&
\left.+\frac{1}{2}\,(\partial_{\mu}\phi)^{2}+\frac{1}{2}\,(\partial_{\mu}X^{\mu})^{2}-m\,\phi\,\partial_{\mu}
X^{\mu} + ie\bar{\psi}\gamma_{5}\psi\phi
+\frac{i}{\sqrt{6}} \, g \,\bar{\psi}\gamma_{5}\psi\,(\partial_{\mu}X^{\mu})\left.\phantom{\frac{1}{2}}\!\!\!\!\right]\right. \, , \label{eq:Reduced_Action}
\end{eqnarray}
where
\begin{equation}
G_{\mu\nu\kappa}=\partial_{\mu}B_{\nu\kappa}+\partial_{\nu}B_{\kappa\mu}+\partial_{\kappa}B_{\mu\nu}\;,\label{eq:field_strength_Kalb-Ramond}
\end{equation}
is the field-strength associated with the Kalb-Ramond field. By considering parity
transformations in $5D$,
we can see that both $\phi$ and $\partial_{\mu}X^{\mu}$ behave as pseudo-scalars in $4D$.
Therefore, the action
(\ref{S4D}) is absolutely parity-invariant in $4D$. The vector field $X_{\mu}$ is the dual of $C_{\mu\nu\kappa}$
\begin{equation}
X^{\mu}:=\frac{1}{\sqrt{6}}\,\epsilon^{\mu\nu\kappa\lambda}C_{\nu\kappa\lambda}\;,
\end{equation}
and, by using the gauge transformation (\ref{eq:Transf_Gauge_C_mu_nu_kappa})
of $C_{\mu\nu\kappa}$, we obtain
\begin{equation}\label{Xtransf}
X^{\mu\,\prime}=X^{\mu}+\frac{1}{2\sqrt{6}}\,\epsilon^{\mu\nu\kappa\lambda}\partial_{\nu}\xi_{\kappa\lambda}\;,
\end{equation}
and hence
\begin{equation}
\partial_{\mu}X^{\mu\,\prime}=\partial_{\mu}X^{\mu}\;,\label{eq:transverse_comp}
\end{equation}
i.e., the vector field $X^{\mu}$ is purely longitudinal. By using
the field equations (\ref{eq:Field_eq_mass-a}-\ref{eq:Field_eq_mass-b}),
this dimensional reduction shows that the bosonic fields in the reduced
action (\ref{eq:Reduced_Action}) acquire a mass $m^{2}$. The field
$\widetilde{H}^{\bar{\mu}}=(\widetilde{G}^{\mu},\widetilde{H}^{4})$
may be split in $\widetilde{H}^{4}= \frac{1}{\sqrt{6}} \partial_{\mu}X^{\mu}$ and
$\widetilde{G}^{\mu}$,
i.e., the dual of $G^{\mu\nu\kappa}$: ${\widetilde{G}^{\mu}=(\chi,\overrightarrow{Y})}$.
The dual of $G^{\mu\nu\kappa}$ is given by
\begin{equation}
\widetilde{G}_{\mu}=\frac{1}{6}\,\epsilon_{\mu\nu\kappa\lambda}G^{\nu\kappa\lambda}\;.\label{eq:Dual de G_4D}
\end{equation}
Therefore, the covariant derivative of (\ref{eq:Reduced_Action})
in four dimensions is
\begin{equation}
D_{\mu}=\partial_{\mu}+ieA_{\mu}+ig\widetilde{G}_{\mu}\;.\label{eq:Derivada_Covariante_4D}
\end{equation}
Here, the 3-form gauge field in 4D is nothing but a longitudinal vector,
because it propagates its longitudinal part and suppresses its transverse
component, as equation (\ref{eq:transverse_comp}) suggests. The light-shining-through-a-wall
experiments (LSW) \citep{Redondo_LSW_2011,betz_LSW_CERN_2013} are
capable of detecting longitudinal radiation \citep{arias_optimizing_2010}.

In connection with the works by Antoniadis \emph{et al}. \citep{antoniadis_axion_2006,antoniadis_anomaly_2008}
and Ringwald \emph{et al}. \citep{arias_optimizing_2010}, what they
consider in 4D as a  pseudo-scalar (Axionic Electrodynamics), turns
out to originate, in our case, from the mixing between the 3-form $(X^\mu)$ and
the $\phi \equiv A^4$ (pseudo-scalar). So, the Antoniadis'
axion is for us a remnant of the 5-dimensional fields in the form of this mixing.

Actually, the papers by Antoniadis \citep{antoniadis_axion_2006,antoniadis_anomaly_2008}
show that our 3-form which appears in 4D must in fact be a pseudo-scalar.
Our vector field, $X^{\mu}$, just propagates the longitudinal part
because this is its gauge invariant component, i.e., this vector field
carries the spin-$\underline{0}$ and the spin-$\underline{1}$ components,
but the gauge symmetry (\ref{Xtransf}) acts to gauge away precisely the spin-$\underline{1}$ piece.

These two new bosons (vector and scalar) that appear simultaneously
in our model can be interpreted, in fact, as \textquotedbl{}two sides
of the same coin\textquotedbl{}. A \textquotedbl{}coin\textquotedbl{}
that is conceived in a 5-dimensional scenario, but, from the point
of view of our 4-dimensional world, leads us to see it as if there
were two separate entities. However, from the point of view of the
five-dimensional bulk, it is only one entity, since the 5 dimensions
provide a unified view of these two fields. In 4D, we see two entities,
the vector and scalar bosons, as a result of dimensional reduction.
Under this unified interpretation, the masses of the \textquotedbl{}two
particles\textquotedbl{} being the same would also suggest an that
there is a common entity the propagates in the bulk between the branes. Further on,
in Section III.C, we shall discuss about the split of this mass degeneracy.

\subsection{Considering the gravitational sector.}

An issue to be investigated concerns the introduction of the gravitational coupling in the action (\ref{acao5D})
to subsequently perform a dimensional reduction to $4D$. For this purpose, we consider the action (\ref{acao5D}),
now in the presence of gravity, to be given by
\begin{eqnarray}\label{acao5DG}
S_{5D}&=&\int d^{5}x \, \sqrt{-g} \left[-\frac{R}{2\kappa^{2}}+\bar{\psi}\;(i\gamma^{\bar{\mu}}{\cal D}_{\bar{\mu}}-m_{f})\;\psi-\frac{1}{4}\,g^{\bar{\mu}\bar{\alpha}}g^{\bar{\nu}\bar{\beta}}
F_{\bar{\mu}\bar{\nu}}F_{\bar{\alpha}\bar{\beta}}
\right. \nonumber \\
&&
\left.
-\frac{1}{8}\,g^{\bar{\mu}\bar{\alpha}}g^{\bar{\nu}\bar{\beta}}g^{\bar{\rho}\bar{\gamma}}g^{\bar{\lambda}\bar{\sigma}}H_{\bar{\mu}\bar{\nu}\bar{\kappa}\bar{\lambda}}
H_{\bar{\alpha}\bar{\beta}\bar{\gamma}\bar{\sigma}}
+\frac{m}{\sqrt{6}}\,
\frac{\epsilon^{\bar{\mu}\bar{\nu}\bar{\kappa}\bar{\lambda}\bar{\rho}}}{\sqrt{-g}}
A_{\bar{\mu}}\partial_{\bar{\nu}}C_{\bar{\kappa}\bar{\lambda}\bar{\rho}}\right] \; ,
\end{eqnarray}
where $\kappa$ is the gravitational coupling (related to Newton constant by $\kappa^{2}=8\pi\,G$), $R$ is the Ricci scalar,
and the covariant derivative, ${\cal D}_{\bar{\mu}}$, acting on the fermions contains the spin connection,
$\Omega_{\bar{\mu}}$, as given below
\begin{equation}\label{eq:Derivada_Covariante_5DG}
{\cal D}_{\bar{\mu}}=\partial_{\bar{\mu}}+ieA_{\bar{\mu}}+ig\widetilde{H}_{\bar{\mu}} + ig^{\prime}\,\Omega_{\bar{\mu}}  \; .
\end{equation}
The coupling of gravity to fermions requires the vielbein formalism, the so-called first-order approach.
Here, we carry out a natural extension of the formalism to five dimensions.
It is known in the literature that the metric of the curved space-time is written as $e_{a}^{\bar{\mu}} \, e_{b}^{\bar{\nu}}\, g_{\bar{\mu}\bar{\nu}}=\eta_{ab}$, in our case, $\eta^{ab}(+,-,-,-,-)$ is the Minkowski metric on the tangent space, and $e_{a}^{\bar{\mu}}$
is the $5$-bein. The spin connection is expanded in the basis of the Lorentz group generators, $\Sigma^{ab}=\frac{i}{4}\left[\gamma^{a},\gamma^{b}\right]$, as $\Omega_{\bar{\mu}}(x)=\frac{1}{2} \, \Sigma^{ab} \, \omega_{\bar{\mu}}^{ab}(x)$, where $a,b=\{0,1,2,3,4\}$ are the frame indices of the Lorentz group. The gamma-matrices, $\gamma^{\bar{\mu}}$, are defined as $\gamma^{\bar{\mu}}=\gamma^{a}e_{\;\;a}^{\bar{\mu}}$ and fulfill the Clifford algebra $\left\{\gamma^{\bar{\mu}},\gamma^{\bar{\nu}} \right\}=e_{a}^{\bar{\mu}} \, e_{b}^{\bar{\nu}} \, \{\gamma^{a},\gamma^{b}\}=2\,g^{\bar{\mu}\bar{\nu}}$.
The components of the spin connection are related to the vielbein and metric as follows :
\begin{eqnarray}
\omega^{ab}_{\bar{\mu}}=\frac{1}{2}\,e^{a}_{\bar{\nu}}\,\partial_{\bar{\mu}}e^{b\bar{\nu}}
+\frac{1}{2}\,e^{a\bar{\nu}}\,e^{b\bar{\sigma}}\,\partial_{\bar{\sigma}}g_{\bar{\mu}\bar{\nu}}
-\frac{1}{2}\,e^{b}_{\bar{\nu}}\,\partial_{\bar{\mu}}e^{a\bar{\nu}}
-\frac{1}{2}\,e^{b\bar{\nu}}\,e^{a\bar{\sigma}}\,\partial_{\bar{\sigma}}g_{\bar{\mu}\bar{\nu}} \; .
\end{eqnarray}
In the sector of the gauge fields, the tensor $F_{\bar{\mu}\bar{\nu}}$ remains unaltered when coupled to the covariant derivative
of the curved space-time. The same is true for the $3$-form $H_{\bar{\mu}\bar{\nu}\bar{\rho}\bar{\lambda}}$. The 2- and 3-forms
$F_{\bar{\mu}\bar{\nu}}$ and $H_{\bar{\mu}\bar{\nu}\bar{\rho}\bar{\lambda}}$, even if defined with the usual derivatives, behave
like tensors and so there is no need to redefine them by replacing the ordinary by the covariant derivatives. Moreover, if the latter
are used to redefine $F$ and $G$, the gauge symmetries for $A_{\bar{\mu}}$ and $C_{\bar{\mu}\bar{\nu}\bar{\rho}}$ would be explicitly
broken if torsion is present. And this is the case, since we have fermions. So, to keep the gauge symmetries, the expressions for $F$
and $H$ are not changed in presence of gravity, and covariance under general coordinate transformations is also guaranteed. To get information
on the excitation spectrum of the gravity sector, we take the linear approximation for the gravitational field :
\begin{eqnarray}
g_{\bar{\mu}\bar{\nu}}(x)=\eta_{\bar{\mu}\bar{\nu}}+\kappa \, h_{\bar{\mu}\bar{\nu}}(x) \; ,
\end{eqnarray}
where we consider just linear terms in the $\kappa$ constant. In so doing, we obtain the action (\ref{acao5DG}) linearized in $5D$ as
\begin{eqnarray}\label{acao5DGh}
S_{5D}&=&\int d^{5}x \,  \left[-\frac{1}{4}\left(\partial_{\bar{\mu}}h_{\bar{\nu}\bar{\rho}}\right)^{2}
+\frac{1}{2}\,\left(\partial_{\bar{\mu}}h^{\bar{\mu}\bar{\nu}}\right)^{2}
+\frac{1}{2}\,{\bar h}\,\partial_{\bar{\mu}}\partial_{\bar{\nu}}h^{\bar{\mu}\bar{\nu}}
+\frac{1}{4}\left(\partial_{\bar{\mu}}{\bar h}\right)^{2}+
\right. \nonumber \\
&&
\left.
+\bar{\psi}\;(i\gamma^{\bar{\mu}}D_{\bar{\mu}}-m_{f})\;\psi-\frac{1}{4}\,
F^{\,2}_{\bar{\mu}\bar{\nu}}
-\frac{1}{8}\,H^{\,2}_{\bar{\mu}\bar{\nu}\bar{\kappa}\bar{\lambda}}
+\frac{m}{\sqrt{6}}\,\epsilon^{\bar{\mu}\bar{\nu}\bar{\kappa}\bar{\lambda}\bar{\rho}}A_{\bar{\mu}}
\partial_{\bar{\nu}}C_{\bar{\kappa}\bar{\lambda}\bar{\rho}}+{\cal O}(\kappa)\right] \; ,
\end{eqnarray}
where $\bar{h}:=h_{\bar{\mu}}^{\;\;\,\bar{\mu}}$ and we have omitted the terms of order
${\cal O}(\kappa)$ that include the gravitational interactions of
the fermions and gauge fields, since they are not important for what
we shall discuss in the sequel. Actually, we wish to keep track of the interference, in
4D, between the degrees of freedom stemming from the gravitational sector and the bosonic
fields in the gauge sector. This is why we include the gravity-fermion interactions in
the ${\cal O}(\kappa)$-term of the action above. In this action, the $h$-Lagrangian is
invariant under the gauge transformation%

\begin{eqnarray}\label{htransf}
h_{\bar{\mu}\bar{\nu}} \longmapsto h_{\bar{\mu}\bar{\nu}}^{\prime}=h_{\bar{\mu}\bar{\nu}}
+\kappa^{-1}\left(\partial_{\bar{\mu}}\xi_{\bar{\nu}}+\partial_{\bar{\nu}}\xi_{\bar{\mu}} \right) \; ,
\end{eqnarray}
where $\xi_{\bar{\mu}}$ is any vector function in $5D$.
Now, we investigate the dimensional reduction to $4D$ in the kinetic terms of the $h$-field by splitting the components $h^{\bar{\mu}\bar{\nu}}=\left\{h^{\mu\nu},h^{\mu4},h^{44} \right\}$, and defining the components $h^{\mu4}:=V^{\mu}$, $h^{44}:=\chi$.
We adopt the previous condition that $\partial_{4}(\mbox{any field})=0$, so the $5D$ action takes the form below in $4D$ :
\begin{eqnarray}\label{acao5DGh2}
S_{4D}&=&\int d^{4}x \,  \left[-\frac{1}{4}\left(\partial_{\mu}h_{\nu\rho}\right)^{2}
+\frac{1}{2}\,\left(\partial_{\mu}h^{\mu\nu}\right)^{2}
+\frac{1}{2} \, h \, \partial_{\mu}\partial_{\nu}h^{\mu\nu}
+ \frac{1}{4} \, \left(\partial_{\mu}h\right)^{2}
\right. \nonumber \\
&&
\left.
-\frac{1}{4}\,\left(\partial_{\mu}V_{\nu}- \partial_{\nu}V_{\mu}\right)^{2}
+\bar{\psi}\;(i\gamma^{\mu}D_{\mu}-m_{f})\;\psi
\right. \nonumber \\
&&
\left.
-\frac{1}{4}\,F_{\mu\nu}^{\;2}+\frac{1}{6}\,G_{\mu\nu\kappa}^{\;2}
-\frac{2\sqrt{2}}{3}\,m\,\epsilon^{\mu\nu\kappa\lambda}A_{\mu}\partial_{\nu}B_{\kappa\lambda}
\right. \nonumber \\
&&
\left.
+\frac{1}{2}\,(\partial_{\mu}\phi)^{2}+\frac{1}{2}\,(\partial_{\mu}X^{\mu})^{2}-\,m \,\phi\, \partial_{\mu}X^{\mu}+{\cal O}(\kappa)\right] \; ,
\end{eqnarray}
where $h:=h_{\mu}^{\;\;\,\mu}$. In this expression, we notice the emergence of a new vector field, $V^{\mu}$, and a mixing term
of a scalar field, $\chi$, with the weak gravitational field $h^{\mu\nu}$. The kinetic term for the $\chi$-field
naturally drops out. It is then reasonable to truncate the $\chi$-field in the reduction, so that
$\chi=0$. The kinect term for the vector field $V^{\mu}$ is
invariant under the gauge transformation, $V_{\mu} \longmapsto V_{\mu}^{\prime}=V_{\mu}+\kappa^{-1}\partial_{\mu}\xi_{4}(x)$; this is
readily checked by making the dimensional reduction in (\ref{htransf}). Therefore, we have obtained an action in $4D$ with three vector fields,
$A^{\mu}, V^{\mu}, X^{\mu}$, in which there is no mass term associated to the $V^{\mu}$-field.



\subsection{The propagators of the $\left\{ A^\mu, B^{\nu \kappa}, X^\alpha, \phi
\right\}-$multiplet.}

\label{sub:Propagators}

The propagators associated with the Lagrangian (\ref{eq:Reduced_Action})
are obtained after the inclusion of the corresponding gauge-fixing terms:
\begin{equation}
\mathcal{L}_{gf}=-\frac{1}{2\alpha}\,(\partial_{\mu}A^{\mu})^{2}-\frac{1}{2\beta}\,
(\partial_{\mu}B^{\mu\nu})^{2} - \frac{1}{4 \xi} \left( \partial_\mu X_\nu - \partial_\nu
X_\mu \right)^2 \; . \label{eq:termo de gauge fixing-2}
\end{equation}
By adding it to the free part of (\ref{eq:Reduced_Action}), we have
$\mathcal{L}_{0}=\mathcal{L}_{04D}+\mathcal{L}_{gf}$; where
\begin{eqnarray}
\mathcal{L}_{04D} & = & \left.\bar{\psi}\;(i\gamma^{\mu}D_{\mu}-m_{f})\;\psi-\frac{1}{4}
\,
F_{\mu\nu}^{\;2}-\frac{1}{2\alpha}\,(\partial_{\mu}A^{\mu})^{2}+\frac{1}{6}\,G_{\mu\nu\kappa}^{\;2}\right.\nonumber \\
 &  & \left.-\frac{1}{2\beta}\,(\partial_{\mu}B^{\mu\nu})^{2}
-\frac{2\sqrt{2}}{3}\,m\,\epsilon^{\mu\nu\kappa\lambda}A_{\mu}\partial_{\nu}B_{\kappa\lambda}\right.\nonumber \\
 &  &
\left.+\frac{1}{2}\,(\partial_{\mu}X^{\mu})^{2}+\frac{1}{2}\,(\partial_{\mu}\phi)^{2}-\,m\,\phi\,\partial_{\mu}X^{\mu}\;.\right.
\end{eqnarray}
In the sector of gauge fields, we cast the Lagrangian into the
form below:
\begin{eqnarray}
\mathcal{L}_{04D} & = & \left.\frac{1}{2}\,A^{\mu} \,
\Box\left(\theta_{\mu\nu}+\frac{1}{\alpha} \, \omega_{\mu\nu}\right)A^{\nu}\right.\nonumber \\
&  & \left.-\frac{1}{2}B^{\mu\nu}\Box\left[\left(P_{b}^{1}\right)_{\mu\nu,\kappa\lambda}
+\frac{1}{2\beta}\left(P_{e}^{1}\right)_{\mu\nu,\kappa\lambda}\right]B^{\kappa\lambda}\right.\nonumber \\
&  & \left.-\frac{1}{2}\phi\Box\phi-\frac{1}{2} \, X^{\mu}\Box\omega_{\mu\nu}X^{\nu}-\frac{\sqrt{2}}{3}\,m\,A^{\mu}S_{\mu\kappa\lambda}B^{\kappa\lambda}\right.\nonumber \\
&  & \left.+\frac{\sqrt{2}}{3}\,m\,B^{\kappa\lambda}S_{\kappa\lambda\mu}A^{\mu}
-\frac{1}{2}\,m\,\phi\,\partial_{\mu}X^{\mu}\right.\nonumber \\
&  & \left.+\frac{1}{2}\,m\,X^{\mu}\partial_{\mu}\phi\;,\right.
\end{eqnarray}
written in terms of the projection operators:
\begin{equation}
\theta_{\mu\nu}+\omega_{\mu\nu}=\eta_{\mu\nu}
\hspace{0.3cm},\hspace{0.3cm}
\omega_{\mu\nu}=\frac{\partial_{\mu}\partial_{\nu}}{\Box}
\end{equation}
\begin{equation}
\left(P_{b}^{1}\right)_{\mu\nu,\kappa\lambda}=\frac{1}{2}\left(\theta_{\mu\kappa}\theta_{\nu\lambda}-\theta_{\mu\lambda}\theta_{\nu\kappa}\right)\,,
\end{equation}
\begin{equation}
\left(P_{e}^{1}\right)_{\mu\nu,\kappa\lambda}=\frac{1}{2}\left(\theta_{\mu\kappa}\omega_{\nu\lambda}-\theta_{\mu\lambda}\omega_{\alpha\kappa}-\theta_{\nu\kappa}\omega_{\mu\lambda}+\theta_{\nu\lambda}\omega_{\mu\kappa}\right)\,,
\end{equation}
\begin{equation}
S_{\mu\nu\kappa}=-m\epsilon_{\mu\nu\kappa\lambda}\partial^{\lambda}\; ,
\end{equation}
which satisfy the relations
\begin{equation}
\left(P_{b}^{1}\right)_{\mu\nu,\kappa\lambda}\left(P_{b}^{1}\right)_{\;\;\;\;,\;\rho\sigma}^{\kappa\lambda}=\left(P_{b}^{1}\right)_{\mu\nu,\rho\sigma}\hspace{0.2cm},
\end{equation}
\begin{equation}
\left(P_{e}^{1}\right)_{\mu\nu,\kappa\lambda}\left(P_{e}^{1}\right)_{\;\;\;\;,\;\rho\sigma}^{\kappa\lambda}=\left(P_{e}^{1}\right)_{\mu\nu,\rho\sigma}\;,
\end{equation}
\begin{equation}
\left(P_{b}^{1}\right)_{\mu\nu,\kappa\lambda}\left(P_{e}^{1}\right)_{\;\;\;\;,\;\rho\sigma}^{\kappa\lambda}=0\hspace{0.2cm},
\end{equation}
\begin{equation}
\left(P_{e}^{1}\right)_{\mu\nu,\kappa\lambda}\left(P_{b}^{1}\right)_{\;\;\;\;,\;\rho\sigma}^{\kappa\lambda}=0\;,
\end{equation}
\begin{equation}
S_{\mu\nu\alpha}S^{\alpha\kappa\lambda}=-2\,\Box\left(P_{b}^{1}\right)_{\mu\nu}^{\;\;\;\;,\;\kappa\lambda}\hspace{0.2cm},
\end{equation}
\begin{equation}
\left(P_{b}^{1}\right)_{\mu\nu,\alpha\beta}S^{\alpha\beta\kappa}=S_{\mu\nu}^{\;\;\;\;\kappa}\;,
\end{equation}
\begin{equation}
S^{\kappa\alpha\beta}(P_{b}^{1})_{\alpha\beta,}^{\;\;\;\;\;\;\mu\nu}=S^{\kappa\mu\nu}\hspace{0.2cm},
\end{equation}
\begin{equation}
\left(P_{e}^{1}\right)_{\mu\nu,\alpha\beta}S^{\alpha\beta\kappa}=0\hspace{0.2cm},
\end{equation}
\begin{equation}
\left(P_{e}^{1}\right)_{\mu\nu,\alpha\beta}S^{\alpha\beta\kappa}=0\hspace{0.2cm},
\end{equation}
\begin{equation}
S_{\;\;\;\alpha\beta}^{\kappa}(P_{e}^{1})^{\alpha\beta,\mu\nu}=0\;.
\end{equation}
It is convenient to rewrite the Lagrangian in matrix form. For this
task, we split the matrix elements as 
\begin{equation}
P_{\mu\nu}=\Box\left(\theta_{\mu\nu}+\frac{1}{\alpha}\,\omega_{\mu\nu}\right)\;,
\end{equation}
\begin{equation}
Q_{\mu\rho\sigma}=-R_{\mu\nu\sigma}=\frac{2\sqrt{2}}{3}\,m\,S_{\mu\rho\sigma}\;,
\end{equation}
\begin{equation}
{\mathbb{S}}_{\kappa\lambda,\rho\sigma}=
-\Box\left[(P_{b}^{1})_{\kappa\lambda,\rho\sigma}+\frac{1}{2\beta}\,(P_{e}^{1})_{\kappa\lambda,\rho\sigma}\right]\;.
\end{equation}

\begin{equation}
W_{\alpha \beta} \equiv - \Box \omega_{\alpha \beta} + \frac{1}{\xi} \Box \theta_{\alpha
\beta}\end{equation}
Let us write, $\mathcal{L}_{0}=\frac{1}{2}\mathbb{N}^{t}\mathbb{M}\mathbb{N}$,
where ${\mathbb{N}^{t}}=\left(\begin{array}{cccc}
A^{\mu} & B_{\kappa\lambda} & X^{\alpha} & \phi\end{array}\right)$ 
and
\begin{equation}
\mathbb{M}=\left(\begin{array}{cccc}
P_{\mu\nu} & R_{\mu\rho\sigma} & 0 & 0\\
Q_{\kappa\lambda\nu} & {\mathbb{S}}_{\kappa\lambda,\rho\sigma} & 0 & 0\\
0 & 0 & W_{\alpha\beta} & m\partial_{\alpha}\\
0 & 0 & -m\partial_{\beta} & -\Box
\end{array}\right)\;.
\end{equation}
After that, we invert the $\mathbb{M}$-matrix to find the
 propagators listed below:
\begin{equation}\label{propPhiPhi}
\left.\langle\phi \, \phi\rangle=\frac{i}{k^{2}-m^{2}}\;,\hspace{0.2cm}\right.
\end{equation}
\begin{equation}
\left.\langle
X_{\mu}X_{\nu}\rangle=\frac{i}{k^{2}-m^{2}}\frac{k_{\mu}k_{\nu}}{k^{2}}
-i \frac{\xi}{k^2} \left( \eta_{\mu \nu} - \frac{k_\mu k_\nu}{k^2} \right)
\:,\hspace{0.2cm}\right.
\end{equation}
\begin{equation}
\left.\langle \phi \, X^{\mu}\rangle = - \langle X^{\mu} \,
\phi\rangle=\frac{m}{k^{2}-m^{2}}\frac{k^{\mu}}{k^{2}}\:,\hspace{0.2cm}\right.
\end{equation}
\begin{equation}\label{PropAB}
\left.\langle A_{\mu}B_{\nu\kappa}\rangle = -\langle B_{\mu\nu}A_{\kappa}\rangle
=\frac{m}{k^{2}-m^{2}}\frac{\epsilon_{\mu\nu\kappa\lambda}k^{\lambda}}{k^{2}}\:,\hspace{0.2cm}\right.
\end{equation}
\begin{eqnarray}
\langle A_{\mu}A_{\nu}\rangle & = & \left.-\frac{i}{k^{2}-m^{2}}\left[\eta_{\mu\nu}+(\alpha-1)
\frac{k_{\mu}k_{\nu}}{k^{2}}\right]\right.\nonumber \\
 &  & \left.+\alpha\,\frac{im^{2}}{k^{2}-m^{2}}\frac{k_{\mu}k_{\nu}}{(k^{2})^{2}}\:,\hspace{0.2cm}\right.
\end{eqnarray}
\begin{eqnarray}\label{PropBB}
\langle B_{\mu\nu}B_{\kappa\lambda}\rangle & = & \left.\frac{i}{k^{2}-m^{2}}\left[\uma_{\mu\nu,\kappa\lambda}+\left(\beta-\frac{1}{2}\right)\mathbb{K}_{\mu\nu,\kappa\lambda}\right]
-\frac{i \beta m^{2}}{k^{2}-m^{2}}
\frac{\mathbb{K}_{\mu\nu,\kappa\lambda}}{k^{2}}\:,\hspace{0.2cm}\right.\end{eqnarray}
where $\mathbb{K}_{\mu\nu,\kappa\lambda}:=\,\eta_{\mu\kappa}\frac{k_{\nu}k_{\lambda}}{k^{2}}-\eta_{\mu\lambda}\frac{k_{\kappa}k_{\nu}}{k^{2}}
-\eta_{\nu\kappa}\frac{k_{\mu}k_{\lambda}}{k^{2}}+\eta_{\nu\lambda}\frac{k_{\mu}k_{\kappa}}{k^{2}}$  , and
$\uma_{\mu\nu,\kappa\lambda}:=\frac{1}{2}\left( \eta_{\mu\kappa} \eta_{\nu\lambda} - \eta_{\mu\lambda} \eta_{\nu\kappa} \right)$ .
\medskip{}
%

In $5D$, a $1$-form gauge potential carries $3$ on-shell degrees of freedom
(d.f.); a 3-form gauge field propagates just 1 on-shell d.f.. Therefore,
we have $4$ physical degrees of freedom in the sector of gauge bosons.
In $4D$, consequently we must have these $4$ d.f. distributed among the fields
we end up with upon dimensional reduction. Considering the propagators
of the $\phi$- and $X^{\mu}$-sectors ($\phi$ comes
from the Maxwell field in $5D$ and behaves as a pseudo-scalar
in $4D$; $X^{\mu}$ comes from the 3-form in $5D$ and is the dual of the
corresponding $3$-form in $4D$, so it does not propagate any on-shell d.f.),
it becomes clear that the gauge sector in $4D$ also carries 4 d.f., as it
should be.  The other 3 d.f. are carried by the mixed $\left\{ A^\mu, B^{\nu
\kappa} \right\}-$system, in such a way that $A^\mu$ propagates 2 d.f., whereas $B^{\nu 
\kappa}$ carries 1 d.f., due to its gauge symmetry; this then means that these two 
fields mixed together describe a single massive and neutral spin-1 gauge particle, which 
we interpret as a sort of paraphoton. Instead of appearing in a mixed $FF-$term 
\cite{Ahlers_et_al}, our paraphoton is the particle associated to the $\left\{ A_\mu - 
B_{\mu \nu} \right\}-$system with a gauge-invariant mass. 

 Before ending this section, we should clarify two aspects. The first point
concerns the massive pseudo-scalar particle described by the $\left\{ \phi, X^\mu
\right\}-$system. The $5D \rightarrow 4D$ reduction clearly shows that the $X^\mu-$field
appears in 4 dimensions only through its divergence. All terms with $X^\mu$ in eq. (43)
exhibit a $ \partial \cdot X$; $X^\mu$ never appears otherwise. This means that we are 
allowed to actually redefine 
a newfield: $s \equiv \partial_\mu X^\mu$, which is then an auxiliary field and can
therefore be eliminated through its classical field equation:
\begin{equation} s - m \phi + \frac{i}{\sqrt{6}} g \bar{\psi} \gamma_5 \psi = 0 \, .
\end{equation}
 Since $s$ is an auxiliary field, it is correct to replace it in the original action 
(43) through 
its algebraic equation above, from which we get the canonical Klein-Gordon action  
$(\frac{1}{2} \partial_\mu \phi \partial^\mu \phi - \frac{1}{2} m^2 \phi^2)$, along with 
a quartic fermionic interaction term, $(\bar{\psi} \gamma_5 \psi)^2$. This confirms that 
the$\left\{ \phi, X^\mu \right\}-$system describes nothing but a massive pseudo-scalar, 
which we
associate to the axion. The $\left\{ A^\mu, B^{\nu \kappa} \right\}-$ system describes
the 3 on-shell d.f. of a neutral massive spin-1 particle. It is however  
mass-degenerate with $\phi$. Nevertheless, the 4-dimensional model does not stand by 
itself. We suggest, but we do not go through that in details here (it is not our goal) 
that $A^\mu$ may couple to a Higgs sector in such a way
that, upon a spontaneous symmetry breaking induced by this Higgs sector, its mass splits 
from the axion mass.

 We take here the Higgs coupling to $A^\mu$ as given by the usual photon-paraphoton 
kinetic 
mixing $\chi-$parameter \cite{Ahlers_et_al}. According to the detailed discussion in the 
paper by Jaeckel and Ringwald \cite{Jaeckel_Ringwald} \cite{Jaeckel_2013} 
\cite{Jaeckel_2010}, $\chi$ ranges between $10^{-16}$ 
and $10^{-4}$, as consideration based on string theory points to. In our case, if, as 
stated above, the Higgs-paraphoton coupling is given by a $\chi-$parameter in the range 
$10^{-16}$ to $10^{-12}$, the axion-paraphoton mass splitting lies in the sub-eV range 
(we remind that $<Higgs> \sim$ 246 GeV), so that the mass degeneracy is lifted. So, in 
our axion-paraphoton model, 
the $\chi-$parameter is also present, but it appears in the Higgs-paraphoton coupling, 
and it is compatible with axion and paraphoton masses both in the sub-eV scale.


\subsection{Spin-dependent Potentials}

In this Section, we study the profiles of the interparticle (non-relativistic, but
spin-and velocity-dependent) potentials when the virtual particles associated to the
fields involved in the propagators above are exchanged. These potentials could be
suitably extended to macroscopic situations if the exchanged mass, $m$, is small enough.
The spin- and velocity-dependent shapes could find some
possible application for a physics tested at the sub-millimetric scale, actually,
$10^{-1}mm$. But, in the case considered here, the mass does not break gauge symmetry, so
that it would be non-trivial to keep track of the influence of the particular mass
mechanism on the form of the interaction particle. We consider the methodology used in
\cite{wilczek_forces_1984} and \cite{dobrescu2006spin}, for which the
potential can be obtained, in the first Born approximation, by performing the Fourier
integral of the amplitude,
\begin{equation}
V(\overrightarrow{r} ,\overrightarrow{v}) = - \int \frac{d^3\overrightarrow{q}}{(2\pi)^3}
\, \, e^{i \overrightarrow{q} \cdot
\,\overrightarrow{r}} \, \mathcal{A} ( \overrightarrow{q} , m \overrightarrow{v} ) \; .
\end{equation}

In the following, we shall use the center-of-mass frame, whose momentum assignments
are fixed as in Figure \ref{fig:overleaf}.

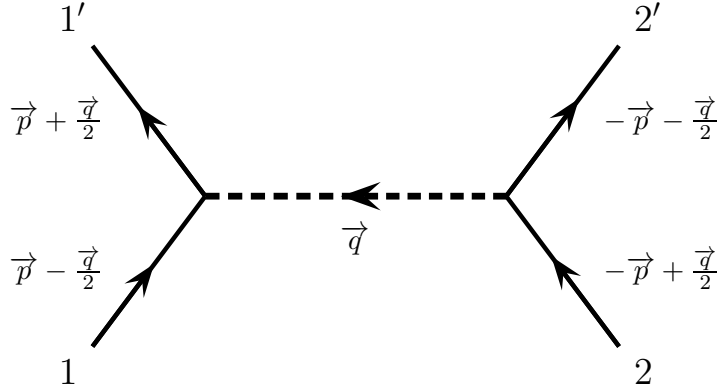
\begin{figure}[!h]
\begin{center}
\newpsobject{showgrid}{psgrid}{subgriddiv=1,griddots=10,gridlabels=6pt}
\begin{pspicture}(0,-0.5)(9,4.6)
\psset{arrowsize=0.2 2}
\psset{unit=1}
%
%
\psline[linestyle=dashed,linecolor=black,linewidth=0.9mm](2.5,2)(4.5,2)
\psline[linestyle=dashed,linecolor=black,linewidth=0.9mm]{<-}(4.3,2)(6.5,2)
\psline[linecolor=black,linewidth=0.6mm]{-}(6.5,2)(8,4)
\psline[linecolor=black,linewidth=0.6mm]{->}(8,0)(7.1,1.2)
\psline[linecolor=black,linewidth=0.6mm]{-}(6.5,2)(8,0)
\psline[linecolor=black,linewidth=0.6mm]{->}(6.5,2)(7.47,3.3)
\psline[linecolor=black,linewidth=0.6mm]{-}(2.5,2)(1,4)
\psline[linecolor=black,linewidth=0.6mm]{->}(1,0)(1.82,1.1)
\psline[linecolor=black,linewidth=0.6mm]{-}(2.5,2)(1,0)
\psline[linecolor=black,linewidth=0.6mm]{->}(2.5,2)(1.6,3.2)
\put(0.55,-0.5){\large$1$}
\put(0.55,4.2){\large$1^{\prime}$}
\put(8.2,-0.5){\large$2$}
\put(8.2,4.2){\large$2^{\prime}$}
%
%
\put(4.3,1.3){$\overrightarrow{q}$}
\put(7.8,2.9){$-\overrightarrow{p}-\frac{\overrightarrow{q}}{2}$}
\put(7.8,0.9){$-\overrightarrow{p}+\frac{\overrightarrow{q}}{2}$}
\put(-0.1,2.9){$\overrightarrow{p}+\frac{\overrightarrow{q}}{2}$}
\put(-0.1,0.9){$\overrightarrow{p}-\frac{\overrightarrow{q}}{2}$}
%
%
%
%
\end{pspicture}
%
%
\caption{\scshape{Momentum assignments in the center-of-mass frame.}}\label{fig:overleaf}
\end{center}
\end{figure}
%


We begin by reviewing a well-known case: two  pseudo-scalar fermionic currents
interacting via the scalar propagator $\langle \phi \phi \rangle$, eq. (\ref{propPhiPhi}).
By applying the Feynman rules, we obtain
\begin{eqnarray}\label{leo_phiphi_1}
i \mathcal{A}^{<\phi \phi>}  =  \bar{u}\left(p + \frac{q}{2}\right)
\left( - e_1 \gamma_5  \right) u\left(p - \frac{q}{2}\right)
\langle \phi \phi \rangle \times
\nonumber \\
\times \, \bar{u}\left(- p - \frac{q}{2}\right) \left( - e_2 \gamma_5  \right)
u\left(- p + \frac{q}{2}\right) \; ,
\end{eqnarray}
which can be rewritten in terms of the pseudo-scalar currents as
\begin{equation}\label{leo_phiphi_2}
\mathcal{A}^{<\phi \phi>}  = e_1 e_2 \, \frac{J_1^{PS}
J_{2}^{PS}}{\overrightarrow{q}^{2} + m^2} \; .
\end{equation}
So, we take the non-relativistic limit for a pseudo-scalar current (see eq.
$(\ref{leo_corrente_PS}) $ in the Appendix) to get:
\begin{equation}\label{leo_phiphi_3}
\mathcal{A}^{<\phi \phi>}  = \frac{e_1 e_2}{4 m_1 m_2} \, \frac{
\left( \overrightarrow{q} \cdot \langle \overrightarrow{\sigma} \rangle_1 \right) \left( \overrightarrow{q} \cdot \langle
\overrightarrow{\sigma} \rangle_2 \right)}{\overrightarrow{q}^{2} + m^2} \, .
\end{equation}
Finally, we carry out the Fourier integral and obtain \cite{wilczek_forces_1984}:
\begin{equation}\label{leo_potencial_PS}
V^{<\phi \phi>}_{PS-PS} = - \frac{e_1 e_2}{4 m_1 m_2} \, \mathcal{V}^{<
\phi \phi >} \; ,
\end{equation}
where we define
\begin{equation}\label{leo_potencial_PS_2}
\mathcal{V}^{< \phi \phi >} =  \Biggl[
\left( 1 + mr \right) \left( \langle \overrightarrow{\sigma} \rangle_1 \cdot
\langle \overrightarrow{\sigma} \rangle_2 \right)
 - \left( 3 + 3mr + m^2 r^2 \right)
\left( \hat{r} \cdot \langle \overrightarrow{\sigma} \rangle_1 \right) \left( \hat{r} \cdot \langle
\overrightarrow{\sigma} \rangle_2 \right)
\Biggr] \, \frac{e^{- mr}}{4 \pi r^3} \; .
\end{equation}
Let us now move on to the next case, where we take the $ \langle A B
\rangle$-propagator of the eq.(\ref{PropAB}). The amplitude is given by
\begin{eqnarray}\label{leo_amplitude_inicio_AB}
i \mathcal{A}^{<AB>} = \bar{u}\left(p + \frac{q}{2}\right) \left(
- i e_1 \, \gamma^\mu \right) u\left(p - \frac{q}{2}\right)   \langle   A_\mu B_{\kappa \lambda}
\rangle \times
\nonumber \\
\times  \bar{u}\left(- p - \frac{q}{2}\right) \left(
 \frac{g_2}{2} \, \gamma_\rho \, \epsilon^{\rho \nu \kappa \lambda } \, q_\nu
\right)  u\left(- p + \frac{q}{2}\right) \, .
\end{eqnarray}
After some algebraic manipulations, it can be rewritten in terms of vector currents:
\begin{equation}\label{leo_amplitude_AB_final}
\mathcal{A}^{<AB>} = -\frac{e_1 g_2 m}{\overrightarrow{q}^{\,\,2} + m^2} \,
\left( J_1^V \right)^\mu \left( J_2^V \right)_\mu  \, .
\end{equation}
If we take the contraction between these currents, eq. $(\ref{leo_contraction})$,
and perform the Fourier integral, we obtain
\begin{equation}\label{leo_potencial_AB}
V^{<AB>} =   e_1 g_2 m \, \delta_1 \delta_2 \, \frac{e^{-mr}}{4 \pi
r} + e_1 g_2 m \, \mathcal{V}_{(2)}^{<AB>} \, ,
\end{equation}
where we have defined
\begin{eqnarray}\label{leo_potencial_AB_2}
\mathcal{V}_{(2)}^{<AB>} &=& \delta_1 \delta_2 \left[ \left( \frac{1}{m_1^2}
+ \frac{1}{m_2^2} \right) \left( \frac{\overrightarrow{p}^{2}}{4} + \frac{m^2}{16} \right)
+ \frac{\overrightarrow{p}^{2}}{m_1 m_2} \right] \, \frac{e^{-mr}}{4 \pi r}
\nonumber \\
&&
-\left\{ \overrightarrow{p} \times \left[ \frac{1}{4} \left(
\frac{\delta_1}{m_2^2} \, \langle \overrightarrow{\sigma} \rangle_2 +
\frac{\delta_2}{m_1^2} \, \langle \overrightarrow{\sigma} \rangle_1  \right)+
\right. \right.
\nonumber \\
&&
\left. \left. +\frac{1}{2} \frac{ \left( \delta_1  \langle \overrightarrow{\sigma} \rangle_2 +
\delta_2  \langle \overrightarrow{\sigma} \rangle_1 \right) }{m_1 m_2}  \right] \right\} \cdot
\hat{r} \left( 1+ mr \right) \frac{e^{-mr}}{4 \pi r^2} +
\nonumber \\
&&+ \left\{ \frac{\left( \langle \overrightarrow{\sigma} \rangle_1 \cdot \langle
\overrightarrow{\sigma} \rangle_2 \right)}{4 m_1 m_2} \left[  1 + mr + m^2 r^2 \right]
\right.
\nonumber \\
&&
\left. -\frac{ \left( \langle \overrightarrow{\sigma} \rangle_1 \cdot \hat{r }\right)
\left( \langle \overrightarrow{\sigma} \rangle_2 \cdot \hat{r }\right)}{4 m_1 m_2}
\left[ 3 + 3 mr + m^2 r^2 \right] \right\} \, \frac{e^{-mr}}{4\pi r^3} \; .
\end{eqnarray}

$\delta_1$ and $\delta_2$ (as explained in details in the Appendix) vanish if the
particle-1 or particle-2 experience spin flip in the interaction.

Thus, we notice that the first term in eq. (\ref{leo_potencial_AB}) behaves like a Yukawa
term, while  $\mathcal{V}_{(2)}^{<AB>}$ is suppressed by a factor of
$\mathcal{O}(v^2/c^2) $.

Finally, we calculate the most involved potential, the one associated with the
$ \langle  B B \rangle $-propagator of eq. (\ref{PropBB}). The amplitude assumes the form
\begin{eqnarray}\label{leo_amplitude_inicio}
i \mathcal{A}^{<BB>} &=& \bar{u}\left(p + \frac{q}{2}\right) \left(
- \frac{g_1}{2} \, \gamma^\rho \, \epsilon_{\rho \xi \mu \nu} \, q^\xi
\right) u\left(p - \frac{q}{2}\right)   \langle  B^{\mu \nu} B_{\kappa \lambda } \rangle
\times\nonumber \\
&&\times \,\, \bar{u}\left(- p - \frac{q}{2}\right) \left(
\frac{g_2}{2} \, \gamma_\alpha \, \epsilon^{\alpha \beta \kappa \lambda } \, q_\beta
\right)  u\left(- p + \frac{q}{2}\right)\nonumber \\
&&= -\frac{g_1 g_2}{4} \,
\epsilon_{\rho \xi \mu \nu } \, \epsilon^{\alpha \beta \kappa \lambda } \, q^{\xi} \,
q_{\beta} \, \left( J^V_1 \right)^\rho  \, \left( J^V_2 \right)_\alpha \,
\langle  B^{\mu \nu} B_{\kappa \lambda } \rangle \; .
\end{eqnarray}
After expressing the product $\epsilon_{\rho \xi \mu \nu } \,
\epsilon^{\alpha \beta \kappa \lambda }$ in terms of Kronecker deltas, this amplitude
simplifies as follows:
\begin{eqnarray}\label{leo_amplitude_meio}
i \mathcal{A}^{<BB>} =  \frac{g_1 g_2}{4} \,\Biggl\{
\left( J_1^V \right)^\mu \left( J_2^V \right)_\mu \left[
2 q^2 \,  \langle  B^{\kappa \lambda } B_{\kappa \lambda } \rangle +4 q_\beta \, \langle
B^{\beta \kappa} B_{\kappa \lambda } \rangle \, q^\lambda \right] +
\nonumber \\
+ 4 q^2 \, \left( J_2^V \right)_\alpha \,
\langle  B^{\alpha \lambda} B_{ \lambda \kappa} \rangle \,
\left( J_1^V \right)^\kappa  +
4  \left( J_2^V \right)_\alpha \, q_\beta \,
\langle  B^{\alpha \beta} B_{ \kappa \lambda} \rangle \,
 \left( J_2^V \right)^\kappa \, q^\lambda \,  \Biggr\} \; .
\end{eqnarray}
Its possible to show, after some lengthy evaluations, that the terms associated with the
operator $\mathbb{K}$, eq. (\ref{PropBB}), do not contribute to the amplitude. Therefore,
the amplitude does not depend on the gauge-fixing terms. Then, we could take
\begin{equation}
\langle  B_{\mu \nu} B_{ \lambda \kappa} \rangle \, = \,
\frac{i}{q^2 - m^2} \, \uma_{\mu \nu , \, \lambda \kappa} \, ,
\end{equation}
which leads to the following result
\begin{equation}\label{leo_amplitude_final}
\mathcal{A}^{<BB>} =  \frac{g_1 g_2}{2} \,
\frac{\overrightarrow{q}^2}{\overrightarrow{q}^2 + m^2} \,
\left( J_1^V \right)^\mu \left( J_2^V \right)_\mu  \, .
\end{equation}
Once again, we use the contraction $(\ref{leo_contraction})$.
The Fourier integral yields  the result:
\begin{equation}\label{leo_potencial_BB}
V^{<BB>} =  \frac{g_1 g_2 m^2}{4} \, \delta_1 \delta_2 \frac{e^{-mr}}{4
\pi r} + \frac{g_1 g_2 m^2}{4} \, \mathcal{V}^{<AB>}_{(2)} \; .
\end{equation}

 This potential has the same functional form and behavior as the one obtained in the
$\langle  A B \rangle$-case, eq.  $(\ref{leo_potencial_AB})$.

 Now, we shall clarify some points. We emphasize that, by adopting the Scherk-Schwarz 
dimension
reduction scheme, such that we decompose $\bar{\mu} = (\mu, 4)$ and assume $\partial_4
(any field) = 0$, we neglect the non-zero Kaluza-Klein modes and, consequently, the 
higher-dimensional
Planck mass does not play any role in our approach. Thus, the four-dimensional physics
should receive only  information on the radius of the extra dimension. This parameter has 
not explicitly
appeared  in the potentials simply because we have used the same notation for the
couplings constants in 5D and 4D. By carrying out a dimensional analysis of the fields 
present in the actions
$S_{5D}$ and $S_{4D}$, we could conclude that, for example, $g^{5D}/\sqrt{L} = g^{4D}$,
where $L = \int dx_4$
stands for the length of the extra dimension (the $x_4$-coordinate). In order to
impose constraints on the radius of extra dimension, we should notice that the coupling
$g$ always appears together with the mass $m$ in the potentials. So, we need to combine 
different
 experiments involving these potentials. We are not following this path here, since this 
is not the scope of the present paper. But, we understand that this point should be the 
object of our attention in a forthcoming work.

 The main inheritances from the five-dimensional physics appear in the vertex 
interactions.  The spin-dependent terms show up only at
$\mathcal{O}(v^2/c^2)$, since the dimensional reduction fixes the vertex interactions as
a result of what we have in 5D. If we study the free Lagrangian directly in 4D, eq.
(\ref{S4D}), we have the freedom to fix the interaction by
means of (pseudo-)vector or (pseudo-)tensor currents. These results are presented in
the work of Ref. \cite{{ferreira2014topologically}}, with spin-dependent terms in first
and second orders in $v/c$.

 If we adopt other dimensional reduction schemes, we  expect to get different
results for the potentials. We also point out a new path: we intend to
calculate the interparticle potential directly in 5D and then take the dimensional 
reduction of 
the $5D-$potential, instead of first reducing from $5D$ to $4D$ to then compute the 
potential
in $4D$. The main reason  is that, in $5D-$Minkowski space, we have two spins, associated 
with the $SO(4) \cong SU(2) \times SU(2)-$little group of Poincare group in $5D$, while 
in 4D we have only one spin, since $SO(3)\cong SU(2)$. 

 We have preliminary results taking into account the 
contribution of this
 new spin in the interparticle potentials. In some situations, there shall appear new
corrections to the monopole-dipole and dipole-dipole potentials, which decrease
with a power-law dependence on the radius of the extra dimension, which now play
a more fundamental role than the renormalized coupling constant. We expect to report on
these results soon.

 In four dimensions, the vector boson $A^\mu$ appears as a massive excitation, so 
that, rather than
a Coulomb-like, we have a Yukawa-type potential. Newton's law is reproduced from the
inspection of the graviton sector, for it comes from the linearisation of $\sqrt{-g} R$ and
no mass parameter appears that endows the gravitation with a mass. 

 If we are to interpret $A^\mu$ as a paraphoton in $4D$, then the mass
parameter $m$ should be constrained by the experiments that fix an upper bound on the
axion mass. The axion-paraphoton splitting is taken care of by the $\chi-$parameter, 
which we propose to govern the Higgs-paraphoton coupling.

\section{Concluding Comments.}

\label{sec:Concluding}

One proposes here to investigate a $5D$ electromagnetic model with a (Abelian)
topological mass term built up in terms of a $1$-form
and a $3$-form gauge potential. Such a description may offer some hint
for modeling the so-called dark energy, due to the presence of the
$\Theta_{\;\;4}^{4}$-component of our energy-momentum tensor that
may correspond to a negative pressure and may then be describing an
expanding system.

 Going over into $4D$, by following the particular dimensional reduction procedure
we have adopted, we identify the emergence of a sector we refer to as an extra
dark sector. It is associated to an excitation that acts as a scalar photon, to which
a scalar magnetic-like field is related. In this scenario, in $4D$, there emerges a 
neutral massive vector boson (mass $m$) along with a neutral pseudo-scalar with the same 
mass. The 3-form that yields a negative contribution to the pressure is responsible (with 
its mixing to the $4D$ Abelian gauge boson inherited from the 5-dimensional 1-form) for 
the appearance of two bosons: a longitudinal vector, that is  an auxiliary field, and 
a massive spin-1 particle, that we interpret as a paraphoton. In our formulation, the 
axion (the pseudo-scalar component identified as $A^4$) and the paraphoton come out 
mass-degenerate, both considered in the sub-eV scale. We propose to couple the 
paraphoton to the electroweak Higgs scalar with the $\chi$-parameter as the Higgs gauge 
coupling, so that the axion and the paraphoton have their degeneracy lifted with a 
splitting also in the sub-eV scale. On the other hand,
the massive scalar may be interpreted as the axion remnant of the
Electrodynamics in 5D considering that the Chern-Simons term (Abelian)
in 5D is defined as $\epsilon^{\bar{\mu}\nu\bar{\lambda}\bar{\kappa}\bar{\mu}\bar{\rho}}A_{\bar{\nu}}
F_{\bar{\lambda}\bar{\kappa}}F_{\bar{\mu}\bar{\rho}}$
and its dimensional reduction to 4D leads to the axionic term type:
$\theta F_{\mu\nu}\widetilde{F}^{\mu\nu}$ where $A^{4}=\theta$ \citep{Redondo_LSW_2011}.

By setting $g=0$ , i.e., by eliminating the non-minimal coupling
described by $\widetilde{G}^{\mu}$ in the covariant derivative, the
field $X^{\mu}$ decouples from the fermions; however, the axionic-like
particle remains coupled, for its coupling is electromagnetic. We
then point out that it is possible to decouple the field $X^{\mu}$,
and, at the same time, to keep the axion coupled with the charged
fermionic matter. We believe that it would be interesting to consider,
from the onset, a Chern-Simons term in five dimensions which would
naturally induce the axionic coupling in $4D$: $\theta F_{\mu\nu}\widetilde{F}^{\mu\nu}$.
The $5D$ Abelian Chern-Simons term is cubic in the gauge field and may
provide a very natural scenario to discuss photon self-interactions
and non-linear effects, with potentially interesting consequences
for the electromagnetic interaction in $4$ dimensions. We shall concentrate
some efforts on this particular issue and we intend to report on that
in a forthcoming paper.

As a final open question, we highlight the study of magnetic monopoles
in a $5$-dimensional scenario, where they become extended one-dimensional
objects (i.e., strings) that appear as the dual of point-like charges.
So, in $5D$, magnetic monopoles have their interaction mediated by the
$2$-form Kalb-Ramond field. As a follow-up of the present work, we shall
be concentrating efforts to pursue an investigation of $5D$ Electrodynamics
in the presence of (extended) magnetic monopoles, so that a $1$-, a
$2$- and a $3$-form should all be present and their effect in connection
with negative pressure and the phenomenon of dark energy in four dimensions
should be reassessed.



\appendix

\section{Currents in the non-relativistic limit}

In this Appendix, we collect the currents and  their contractions. We consider the
same conventions and notations as in Ref. \cite{ferreira2014topologically}. In the
non-relativistic limit, the solution to the Dirac equation, with positive energy,
is given by \cite{ryder1996quantum}:
\begin{equation}
u(p) \approx
\xi \left( \begin{array}{c} 1 \\ \frac{\overrightarrow{\sigma} \cdot
\overrightarrow{p}}{2m} \,
\end{array} \right) \, .
\end{equation}

We take $\xi'$ for the Dirac conjugate $\bar{u}(p)$. The pseudo-scalar
current (PS), following the parametrization for the first vertex of Figure
(\ref{fig:overleaf}), can be written as
\begin{equation}\label{leo_corrente_PS}
J_{1}^{PS} = \bar{u}\left(p + \frac{q}{2}\right) \, i \gamma_5 \, u\left(p - \frac{q}{2}\right) =
- \frac{i}{2 m_1} \, \overrightarrow{q} \cdot \langle \overrightarrow{\sigma}\rangle_1 \, ,
\end{equation}
where we use
$\langle \sigma_i \rangle_1 := \xi'^\dagger \, \sigma_i \, \xi$
to denote the expectation value of the spin matrix, $\sigma_i$, of the the particle one.

For the vector current $(V)$,
\begin{equation}
\left(J_{1}^{V}\right)^{\mu} := \bar{u}\left(p + \frac{q}{2}\right) \, \gamma^\mu \, u\left(p - \frac{q}{2}\right)
\, ;\end{equation}
 the $\mu = 0$-component yields:
\begin{equation}\label{leo_corrente_V_1}
\bar{u}\left(p + \frac{q}{2}\right) \, \gamma^0 \, u\left(p - \frac{q}{2}\right) =  \delta_1 +
\frac{\delta_1}{2 m_1^2} \,  \left( \overrightarrow{p}^{\,2} - \frac{\overrightarrow{q}^{\,2}}{8} \right) +
\frac{i}{4 m_1^2} \, \left( \overrightarrow{q} \times \overrightarrow{p} \right) \cdot  \langle \overrightarrow{\sigma} \rangle_1
\end{equation}
where $\delta_1 := \xi'^{\dagger} \xi $, with $\delta_1 = 0$ if the particle-1 changes
the spin orientation; otherwise $\delta_1 = 1$. The same is true for $\delta_2$.

For the space component, $\mu = i$,  we have:
\begin{equation}\label{leo_corrente_V_2}
\bar{u}\left(p + \frac{q}{2}\right) \, \overrightarrow{\gamma} \, u\left(p - \frac{q}{2}\right) =
\frac{\overrightarrow{p}}{m_1} \, \delta_1 - \frac{i}{2m_1} \, \overrightarrow{q} \times
\langle \overrightarrow{\sigma} \rangle_1  \, .
\end{equation}

The second current, associated with the particle-2 or second vertex of Figure
$\ref{fig:overleaf}$ can be obtained by taking $q \rightarrow - q $, $ p \rightarrow -
p$ and by exchanging the labels $1\rightarrow 2$.

Finally, we present the result for the contraction of vector currents, neglecting terms
of the order $\mathcal{O}(v^3/c^3)$,
\begin{eqnarray}\label{leo_contraction}
\left( J_1^V \right)^\mu \left( J_2^V \right)_\mu  \approx \delta_1 \delta_2 +
\delta_1 \delta_2 \left[ \left( \frac{1}{m_1^2} + \frac{1}{m_2^2} \right)
\left( \frac{\overrightarrow{p}^{\,2}}{4} - \frac{\overrightarrow{q}^{2}}{16} \right)
+ \frac{\overrightarrow{p}^{2}}{m_1 m_2}\right] +
\nonumber \\
+ \left( \overrightarrow{q} \times \overrightarrow{p} \right) \cdot \left[ \frac{i}{4} \, \left(
\frac{\delta_1}{m_2^2} \, \langle \overrightarrow{\sigma} \rangle_2 +
\frac{\delta_2}{m_1^2} \, \langle \overrightarrow{\sigma} \rangle_1  \right) +
\frac{i}{2} \frac{1}{m_1 m_2} \left( \delta_1  \langle \overrightarrow{\sigma} \rangle_2 +
\delta_2  \langle \overrightarrow{\sigma} \rangle_1 \right) \right]
\nonumber \\
- \frac{1}{4} \frac{1}{m_1 m_2} \Bigl\{ \overrightarrow{q}^{2} \left(
\langle \overrightarrow{\sigma} \rangle_1 \cdot \langle \overrightarrow{\sigma} \rangle_2 \right) -
\left( \overrightarrow{q} \cdot \langle \overrightarrow{\sigma} \rangle_1 \right)
\left( \overrightarrow{q} \cdot \langle \overrightarrow{\sigma} \rangle_2 \right) \Bigr\} \, .
\end{eqnarray}

\begin{acknowledgments} 

 J. Jaeckel and P.C. Malta are acknowledged for fruitful exchange of ideas and for 
pointing out relevant references. The authors express their gratitude to the agencies 
\emph{\textquotedbl{}Conselho
Nacional de Desenvolvimento Cient\'{i}fico e Tecnol\'ogico\textquotedbl{}}
(CNPq-Brazil) and \emph{\textquotedbl{}Funda\c{c}\~ao Carlos Chagas
Filho de Amparo \`a Pesquisa do Estado do Rio de Janeiro\textquotedbl{}}
(FAPERJ) for the financial support.

\end{acknowledgments} 

 \bibliographystyle{apsrev4-1}
\bibliography{referencias_artigo}

\end{document}